\documentclass{article}

\usepackage{arxiv}

\usepackage{url} %this package should fix any errors with URLs in refs.
\usepackage{lineno} %add line numbers
\usepackage{soul} %text decoration features
\usepackage{amsfonts,amssymb} %fonts and mathmatical symbols
\usepackage{xcolor} %color
\usepackage{siunitx} %format numerical stuff
\usepackage{physics} % better integrals and such
\usepackage{graphicx}

\usepackage{fancyhdr}
\graphicspath{ {./figures/} }

\usepackage{isotope}
\newcommand{\coo}{\ensuremath{\isotope{CO}_2}}

\usepackage{authblk}

\usepackage[acronym,nopostdot,nonumberlist,shortcuts,]{glossaries}

\newacronym{ams}{AMS}{Annual Maxima Series}
\newacronym{cc}{C-C}{Clausius-Clapeyron}
\newacronym{cdf}{CDF}{Cumulative Distribution Function}
\newacronym{crps}{CRPS}{Continuous Ranked Probability Score}
\newacronym{enso}{ENSO}{El Ni\~{n}o--Southern Oscillation}
\newacronym{gcm}{GCM}{General Circulation Model}
\newacronym{gev}{GEV}{Generalized Extreme Value}
\newacronym{ghcn}{GHCN}{Global Historical Climatology Network}
\newacronym{gp}{GP}{Gaussian Process}
\newacronym{idf}{IDF}{Intensity-Duration-Frequency}
\newacronym{logs}{LogS}{Logarithmic Score}
\newacronym{map}{MAP}{Maximum a posteriori}
\newacronym{mcmc}{MCMC}{Markov Chain Monte Carlo}
\newacronym{mle}{MLE}{Maximum Likelihood Estimation}
\newacronym[]{noaa}{NOAA}{the National Oceanographic and Atmospheric Administration}
\newacronym{nuts}{NUTS}{No-U-Turn-Sampler}
\newacronym{qs}{QS}{Quantile Score}
\newacronym{rfa}{RFA}{Regional Frequency Analysis}
\newacronym{roi}{ROI}{region-of-influence}
\newacronym{rcp}{RCP}{Representative Concentration Pathway}

\usepackage[authoryear]{natbib}
\usepackage{cleveref}

\title{Bayesian Spatiotemporal Nonstationary Model Quantifies Robust Increases in Daily Extreme Rainfall Across the Western Gulf Coast}

\author[1]{\textbf{Yuchen Lu}*\thanks{Corresponding author}}
\author[2]{\textbf{Benjamin Seiyon Lee}}
\author[1,3]{\textbf{James Doss-Gollin}}

\affil[1]{Department of Civil and Environmental Engineering, Rice University, Houston, TX 77005}
\affil[2]{Department of Statistics, George Mason University, Fairfax, Virginia 22030}
\affil[3]{Ken Kennedy Institute, Rice University, Houston, TX 77005}

\affil[*]{yl238@rice.edu}

\begin{document}
\maketitle
\begin{abstract}
Precipitation exceedance probabilities are widely used in engineering design, risk assessment, and floodplain management.
While common approaches like \acrshort{noaa} Atlas 14 assume that extreme precipitation characteristics are stationary over time, this assumption may underestimate current and future hazards due to anthropogenic climate change.
However, the incorporation of nonstationarity in statistical modeling of extreme precipitation has faced practical challenges which have restricted its applications.
In particular, random sampling variability challenges the reliable estimation of trends and parameters, especially when observational records are limited.
To address this methodological gap, we propose the Spatially Varying Covariates Model, a hierarchical Bayesian spatial framework that integrates nonstationarity and regionalization for robust frequency analysis of extreme precipitation.
This model draws from extreme value theory, spatial statistics, and Bayesian statistics, and is validated through cross-validation and multiple performance metrics.
Applying this framework to a case study of daily rainfall in the Western Gulf Coast, we identify robustly increasing trends in extreme precipitation intensity and variability throughout the study area, with notable spatial heterogeneity.
This flexible model accommodates stations with varying observation records, yields smooth return level estimates, and can be straightforwardly adapted to analysis of precipitation frequencies at different durations and for other regions.
\end{abstract}

%% ------------------------------------------------------------------------ %%
%
%  INTRODUCTION
%
%% ------------------------------------------------------------------------ %%

\section{Introduction} \label{sec:introduction}

Extreme precipitation poses risks to ecosystems, economies, and communities \citep{lall_ncawater:2018}.
These risks are exacerbated by global warming and urbanization \citep{donat_moreextreme:2016, tedesco_exposure:2020, merz_review:2014}.
Notable recent examples illustrate the potential impacts of extreme precipitation.
For example, in August \num{2017}, Hurricane Harvey brought record-breaking rainfall that led to catastrophic flooding and damaged more than \num{150000} homes in Southeast Texas \citep{oldenborgh_attribution:2017, russell_gmxsst:2020}.
Similarly in July \num{2021}, an extreme rainfall event struck Henan Province in China, resulting in \SI{7.95}{inch} rainfall in an hour, a record in Zhengzhou, leading to hundreds of fatalities and billions of dollars in direct losses \citep{sun_henan2021:2023, luo_zhengzhou2021:2023}.
Moreover, less extreme localized heavy rainfall events can also cause localized flooding, disrupt transportation networks, and damage infrastructure, with large cumulative impacts in many urban areas \citep{moftakhari_nuisanceflooding:2018, rosenzweig_pluvial:2018, lopez-cantu_standards:2018}

To manage these risks, estimates of precipitation exceedance probabilities are widely used in engineering design, floodplain management, and policy analysis.
These estimates are typically disseminated as \gls{idf} curves, which describe the relationship between intensity, duration, and the probability of exceedance for rainfall at a single point in space \citep{martel_climatechange&IDF:2021, gu_extremechina:2022}.
For instance, \gls{idf} curves form the basis of synthetic design storms, which serve as inputs to hydrologic models for developing floodplain maps \citep{fema_hydrologic:2019}.
\Gls{idf} curves are also used to manage less extreme floods, for example to size pipes used for site-scale stormwater management \citep{cook_stormwater:2020, sharma_stormwater:2021}.
In the United States, \gls{noaa} develops Atlas 14, a widely used set of \gls{idf} curves interpolated across space, and similar efforts exist in other regions \citep{silva_can_idf:2021, kourtis_IDFchange:2022, haruna_comparison:2023}.

A key assumption underpinning most \gls{idf} curves in use today is that extreme precipitation characteristics are stationary over time, which means that the probability distribution of extreme precipitation does not change over time \citep{perica_atlas14southeastern:2013, perica_atlas14_texas:2018}.
However, anthropogenic climate change and natural climate variability drive changes in the frequency, magnitude and spatiotemporal characteristics of heavy precipitation on multiple timescales \citep{milly_stationary:2008, pendergrass_variability:2017, seneviratne_IPCC:2021, sun_change:2021}.
Climate change contributes to extreme precipitation through both thermodynamic and dynamic changes \citep{ogorman_extremes:2015}.
First-order thermodynamic responses are well-understood: following the \gls{cc} relation, a warmer atmosphere holds more water, which can lead to increased rainfall.
Dynamical changes are related to changes in weather patterns, and are more uncertain given the greater differences across different mechanisms and regions.
These dynamical changes could amplify precipitation beyond what the \gls{cc} scaling rate predict, or could lead to less extreme precipitation in some regions \citep{oldenborgh_attribution:2017, risser_Harvey:2017}.
Furthermore, changes in land use and land cover could have an impact on not only the rainfall-runoff relationships but also the fine-scale characteristics of extreme precipitation \citep{pielkesr_landimpact:2007, mahmood_landcover:2014, aich_landuse:2016, zhang_harvey:2018, sui_urban_prcp:2024}.

Recognition of these factors has motivated numerous calls to advance beyond the stationarity assumption \citep{milly_stationary:2008, cheng_nonstationary:2014}.
Yet credibly and reliably incorporating nonstationarity into extreme precipitation probabilities remains a methodological challenge.
Estimates of trends must come either from \glspl{gcm} or from observations, and both approaches have fundamental limitations.
Methods based on \glspl{gcm} must address challenges such as the coarse spatial and temporal resolution of \gls{gcm} outputs, which require downscaling and bias correction approaches that can introduce substantial error and uncertainty \citep{lafferty_BCuncertain:2023, ehret_biascorrection:2012, farnham_credibly:2018, cannon_GCMbc:2015}.
More critically, dynamical biases, whether stemming from inadequate sampling of global dynamics such as \gls{enso} \citep{feng_enso_diversity:2019}, inaccuracies in simulating regional and local processes that govern processes such as tropical cyclone formation, growth, and steering \citep{sobel_biases:2023}, or meridional biases in jet stream location \citep{farnham_credibly:2018}, substantially affect local hazard estimates.
These biases are not easily addressed through post-processing and often result in high uncertainty both within and across models and methods \citep{lafferty_BCuncertain:2023, dittes_uncertainty:2018}.

While observation-based approaches to trend estimation address many of these problems by relying on actual observations rather than on imperfect global models, short observational records and the rare nature of extreme events pose long-standing and well-recognized statistical challenges relating to sampling variability \citep{merz_review:2014}.
Various methodologies attempt to navigate these challenges in different ways, each with its strengths and limitations.
One approach is to use a stationary model within a moving window, which allows for trends to be estimated implicitly without explicit parameterization.
However, this approach exacerbates the problem of estimating the probability of rare events from short records \citep{doss-gollin_robustadaptation:2019}.
For example, \cite{fagnant_spatiotemporal:2020} apply a moving window estimator to rainfall probabilities on the Gulf Coast and find physically implausible discrepancies in trends and return level estimates between nearby locations as well as high volatility in return level estimates at a given location.
Another approach is to model nonstationarity by conditioning the parameters of a statistical distribution on time-varying covariates, ideally credibly simulated by \glspl{gcm} and having a physical relationship with local precipitation \citep{oldenborgh_attribution:2017, russell_gmxsst:2020}.
While commonly applied to hydrologic analysis \citep{schlef_idf:2023}, this ``process-informed'' nonstationary framework increases the number of parameters to estimate and thus exacerbates the problem of sampling variability intrinsic to precipitation frequency analysis \citep{montanari_immortal:2014, serinaldi_undead:2015}.

To mitigate sampling variability between nearby stations, information can be pooled across space in a technique known as regionalization.
In essence, regionalization methods assume that rainfall observed at one location is representative of rainfall at nearby locations, and thus can compensate for stations with short records by combining records from multiple gauges.
Physically, this can be interpreted as a form of spatial averaging analagous to stochastic storm transposition \citep{wright_transposition:2020} and can reduce sampling variability related to the random spatial structure of precipitation in a given storm.
Many regionalization methods have been proposed.
Notably, \acrlong{rfa} \citep{hosking_rfa:1997} aggregates data within the defined homogeneous regions that share similar climate and hydrologic characteristics.
Similarly, \acrlong{roi} approaches \citep{burn_regionofinfluence:1990} dynamically group each station with nearby stations based on similarity criteria, resulting in smooth transitions across regional boundaries.
While these and other regionalization methods have been widely used in the development of \gls{idf} curves \citep{perica_atlas14_texas:2018, fowler_rfauk:2003}, many common implementations rely on $L$-moments, which are not well-suited for nonstationary analysis \citep{salas_review:2018}.
Other studies have employed a wide variety of alternatives, including spatial Bayesian hierarchical models \citep{davison_spatialextremes:2012, cressie_spatiotemporal:2011, lee_picar:2022, wikle_hierarchical:2019, cooley_spatial:2007}, max-stable processes \citep{stephenson_maxstable:2016}, and semi-Bayesian methods \citep{ossandon_semibayesian:2021} to pool information across space.

\subsection{Objectives} \label{sec:objectives}

To integrate the concepts of nonstationarity and regionalization into a coherent statistical framework for precipitation frequency analysis, we propose the Spatially Varying Covariates Model.
This hierarchical Bayesian spatial framework can be applied to stations with varying measurement periods and yields robust extreme precipitation probability estimates.
We apply this framework to daily extreme precipitation on the Western Gulf Coast, but it can be readily applied to other hazards and regions.
Our objective is to shed light on three key research questions:
\begin{enumerate}
    \item Can a hierarchical Bayesian space-time model accurately estimate the parameters of nonstationary extreme value distributions at gauged and ungauged locations?
    \item How do extreme precipitation probabilities vary across space and time?
    \item How do rainfall probability estimates from stationary models compare to those from nonstationary models?
\end{enumerate}
We proceed as follows.
\Cref{sec:methodology} provides methodological details on our model and a case study of daily rainfall probabilities on the Western Gulf Coast.
\Cref{sec:results} presents results for our Gulf Coast case study.
A discussion of our findings in \cref{sec:discussions} is followed by a brief summary in \cref{sec:conclusions}.

%% ------------------------------------------------------------------------ %%
%
%  Methodology
%
%% ------------------------------------------------------------------------ %%

\section{Methodology} \label{sec:methodology}

In this section, we iteratively introduce components of our proposed Spatially Varying Covariates Model.
We begin by describing fundamental concepts in \cref{sec:methodology-background} and introduce our case study in \cref{sec:methods-case-study}.
Next, we outline the three models we use for comparison in \cref{sec:proposed-framework}, with the full Spatially Varying Covariates Model described in \cref{sec:methods-spatially-varying-covariate}.
Finally, we describe the validation and estimation methods in \cref{sec:methods-validation} and \cref{sec:methods-estimation}, respectively.

\subsection{Conceptual Framework} \label{sec:methodology-background}

Here, we present key theoretical building blocks that we integrate to form our Spatially Varying Covariates Model.

\subsubsection{Stationary Model} \label{sec:background-stationary-gev}

Precipitation frequency analysis relies on statistical extreme value theory to estimate the probability of extreme precipitation events \citep{coles_extremes:2001}.
In this work we model block maxima using the \gls{gev} distribution, based on theoretical justification and extensive testing \citep{coles_extremes:2001, cooley_spatial:2007, mishra_extremechange:2010, perica_atlas14_texas:2018}, but our approach could be extended to other distributions or to peaks over threshold analysis.

A stationary extreme value model assumes that the data generating process is invariant over time.
Under the stationary \gls{gev} model, the probability distribution of annual maximum precipitation $y$ in year $t$ at station $s$ is given by:
\begin{equation}
    \label{eqn:background-stationary}
    y(s, t) \sim \mathrm{GEV}(\mu(s), \sigma(s), \xi(s))
\end{equation}
where $\mu$ is the location parameter, $\sigma$ (positive) is the scale parameter, and $\xi$ is the shape parameter.
Often a single station is analyzed, in which case the $s$ may be dropped.

\subsubsection{Nonstationary \acrshort{gev} Model} \label{sec:background-nonstationary-gev}

Nonstationary extreme value statistics extend the traditional stationary framework by allowing parameters to vary in time for each station.
Most generally, we can write the nonstationary model as:
\begin{equation}
    \label{eqn:background-nonstationary}
    y(s, t) \sim \mathrm{GEV}(\mu(s, t), \sigma(s, t), \xi(s, t))
\end{equation}
There are many ways to model nonstationarity \citep{salas_review:2018, schlef_idf:2023}.
Here, we consider the process-informed approach, which conditions the \gls{gev} parameters on climate indices:
\begin{equation}
    \label{eqn:background-nonstationary-params}
    \theta(s,t) = \alpha (s) + \sum_{i} \beta_i(s)x_i(t)
\end{equation}
for a generic nonstationary \gls{gev} parameter $\theta(s,t)$, $\alpha(s)$ represents the corresponding regression intercept, $x_i(t)$ is a time-varying climate covariate, and $\beta_i(s)$ is the associated regression coefficient.
For \gls{gev} models, either the location, scale, or both parameters can be modeled as nonstationary as in \cref{eqn:background-nonstationary-params}.
Suitable covariates should be chosen based on a physical understanding of the data-generating process such that they are causally related to extreme precipitation in a given region and.
If projections are desired, these covariates should also be credibly simulated by global climate models \citep{farnham_credibly:2018}.
As with the stationary framework, the $s$ notation is often dropped when analyzing a single station.

\subsubsection{Gaussian Process} \label{sec:background-GP}

Previous research employing the process-informed framework (\cref{sec:background-nonstationary-gev}) to estimate nonstationary extreme precipitation probabilities has found large estimation uncertainty \citep{cheng_nonstationary:2014, cooley_spatialhierarchical:2010} and implausibly large spatial variability when analyzing single stations \citep{fagnant_spatiotemporal:2020}.
In widely used \gls{idf} curves such as \gls{noaa} Atlas 14, this is addressed though smoothing as a post-processing step \citep{perica_atlas14_texas:2018}.

To mitigate the parametric uncertainty stemming from the increasing number of parameters and limited observation data, regionalization techniques that pool information across locations are commonly applied.
Methods such as \acrlong{rfa} and \acrlong{roi} rely on defining homogeneous regions or grouping stations based on similarity, and typically use the $L$-moments method.
Alternatively, some research use spatial Bayesian hierarchical frameworks, treating parameters as latent functions of space \citep{lee_picar:2022, ulrich_idf:2020, cooley_spatial:2007}.
These frameworks eliminate the need for multi-step processes like defining regions and offer flexibility to incorporate covariates.

To capture spatial dependencies, we incorporate a latent \gls{gp} layer to model the spatially varying distribution parameters.
\Glspl{gp} are non-parametric models, providing full probability distributions not limited to point estimates \citep{rasmussen_gp4ml:2006}.
An advantage of this method is that we learn the spatial dependencies directly from the data, avoiding the need to define homogeneous clusters \textit{a priori}.
Using this method, the spatially varying distribution parameters can be estimated at all gauged stations and interpolated to ungauged locations, enabling the generation of gridded estimates from gauge points.
\Glspl{gp} model the joint distribution of parameters at multiple locations using a multivariate normal distribution with a mean function and a kernel function:
\begin{equation}
    \label{eqn:background-GP-joint}
    f(x) \sim \mathcal{GP}(m(x), K(x,x'))
\end{equation}
where $m(x)$ is the mean function and $K(x,x')$ is the kernel function that characterizes the similarity between two locations.
We employ a constant mean function $m(x) = c$.
Following some experimentation, we employ the exponential kernel, which has been widely used in previous research \citep{cooley_spatial:2007, ossandon_semibayesian:2021} and has desirable convergence properties \citep{rasmussen_gp4ml:2006, cooley_spatial:2007}.
The exponential kernel is given by:
\begin{equation}
    \label{eqn:background-GP-kernel}
    K(x_i,x_j) = \alpha^2 \exp \left( \frac{ -|x_i-x_j| }{\rho} \right)
\end{equation}
where $|x_i-x_j|$ is the euclidean distance between two locations, $\alpha > 0$ is the kernel variance, and $\rho > 0$ is the kernel length parameter.
Alternative kernels may be considered in future work.

The posterior distributions of parameters are correlated, such that changes in one parameter would influence the values of others.
In this research, we adopt univariate \glspl{gp} to model the spatially varying parameters, which generates stable estimates without specifically enforcing correlations.
Future work could explore methods like co-regionalization to account for this correlation \citep{gelfand_spatial:2005, schmidt_coregionalization:2003}.

\subsection{Case Study} \label{sec:methods-case-study}

Our framework can be applied to extreme value analysis of any climate variable, but we focus on daily rainfall in the Western Gulf Coast region.
Adjacent to the Gulf of Mexico, this area has experienced a substantial number of heavy rainfall and flooding events in the recent past.
For instance, Harris County experienced the Memorial Day flood in \num{2015}, when Brays Bayou received approximately \SI{11}{inch} of rain in \num{12} hours \citep{bass_memorial:2017}.
Similarly, Hurricane Harvey in \num{2017} dropped more than \SI{47.2}{inch} rainfall over and around Houston in \num{6} days \citep{zhang_harvey:2018}.
New Orleans has also endured a number of extreme rainfall events, such as Hurricane Katrina, which brought \SI{14}{inch} inches of rain in a single day in August \num{2005} \citep{kates_nolakatrina:2006}.
In August \num{2016}, another intense precipitation event led to a total of \SI{25.5}{inch} over a \num{3}-day period, resulting in severe flooding that impacted around \num{60600} homes \citep{vanderwiel_nola2016:2017}.

We extract \gls{ams} from the \gls{ghcn} daily dataset for selected stations.
Our model accommodates stations with both short and long observation records, but here we retain stations with more than \num{30} years of available data and at least \num{99}$\%$ data coverage each year.
This selection leaves us with \num{181} stations with observations ranging from \num{1889} to \num{2022}.
To ensure consistency with \gls{noaa} Atlas 14 and to facilitate comparison, we converted the constrained \num{1}-day records to unconstrained \num{24}-hour records by multiplying the raw intensity by $1.11$ as in \gls{noaa} Atlas 14 \citep{perica_atlas14_texas:2018}.

\begin{figure}[h]
    \centering
    \includegraphics[width=0.7\textwidth]{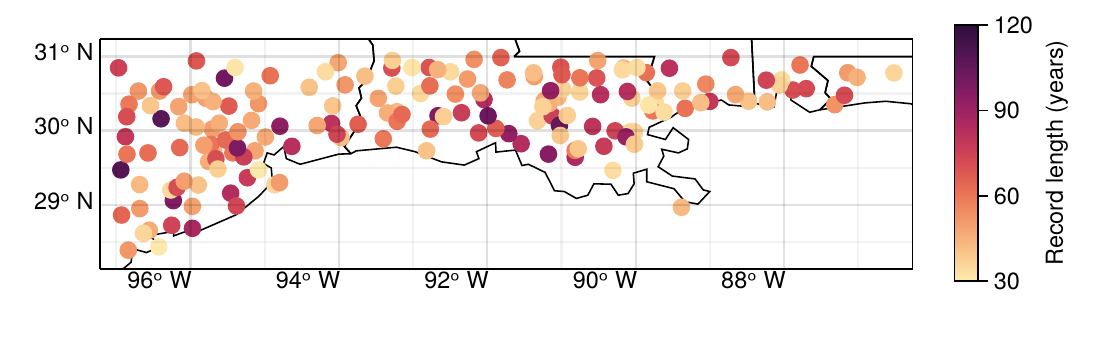}
    \caption{
        Figure shows the \num{181} selected stations from \gls{ghcn} daily dataset.
        The colors indicate the number of available years, ranging from a minimum of \num{30} years to a maximum of approximately \num{120} years.
    }\label{fig:case-study-stations}
\end{figure}

\subsection{Proposed Framework} \label{sec:proposed-framework}

We propose the Spatially Varying Covariate Model (\cref{sec:methods-spatially-varying-covariate}), a fully probabilistic hierarchical spatial Bayesian framework that integrates nonstationarity and regionalization, to analyze the temporal trends of daily extreme precipitation in the Western Gulf Coast.
This model integrates the concepts outlined in \cref{sec:background-stationary-gev,sec:background-nonstationary-gev,sec:background-GP}.
To compare the performance of our proposed framework, we also apply two benchmark models to analyze the study area: the Pooled Stationary Model (\cref{sec:methods-pooled-stationary}) and the Nonpooled Nonstatioanry Model (\cref{sec:methods-nonpooled-nonstationary}).

\subsubsection{Pooled Stationary Model} \label{sec:methods-pooled-stationary}

The \textbf{Pooled Stationary Model} incorporates regionalization within a stationary framework, conceptually similar to the method used in \gls{noaa} Atlas 14 \citep{perica_atlas14southeastern:2013, perica_atlas14_texas:2018}.
In this model, we assume that the \gls{gev} parameters are constant over time (\cref{eqn:background-stationary}), with the location and scale parameters spatially pooled using univariate \glspl{gp}.
To ensure positivity, scale parameters are log-transformed.
Given the relatively small size of the chosen study area, which primarily consists of coastal regions with little elevation variation, shape parameters are set as constant for all stations in this study \citep{cooley_spatial:2007, apputhurai_spatiotemporal:2013}.
For a larger study area encompassing diverse climate and topographic conditions, varying shape parameters is likely to be more advantageous.

For the \glspl{gp} representing the spatially varying parameters (namely, the \gls{gev} location and scale parameters), we employ \emph{Inverse Gamma} and \emph{Gamma} distributions as the prior distributions for kernel variance ($\alpha_k$) and length ($\rho_k$) parameters respectively, as they are strictly positive, as recommended by the Stan manual \citep{carpenter_stan:2017, stan_userguide:2022}.
Priors of these two kernel parameters for both location and scale parameters follow \Cref{eqn:GP-kernel-variance-prior,eqn:GP-kernel-length-prior}.
For simplicity, we fix the \gls{gp} mean ($m_k$) at \num{0}, which does not substantially impact the final estimates.
The prior for the shape parameter is informed by existing literature \citep{lima_GEVflood:2016} and preliminary single station analysis, with absolute values typically below \num{0.5}.
Additionally, in anticipation of a positively skewed distribution for extreme precipitation events, we enforce a positive shape parameter \citep{blanchet_invariant:2016, anshsrivastava_raedaridf:2023}.
The model is thus given as follows, and the detailed priors are shown in \cref{sec:methods-spatially-varying-covariate}.
\begin{align}
    y(s,t)    & \sim \mathrm{GEV}(\mu(s), \sigma(s), \xi) \label{eqn:pooled-stationary-model} \\
    \mu(s)    & \sim \mathrm{GP}(0, K_\mu(s,s')) \label{eqn:pooled-stationary-location}           \\
    \sigma(s) & \sim \mathrm{GP}(0, K_\sigma(s,s')) \label{eqn:pooled-stationary-scale}
\end{align}

\subsubsection{Nonpooled Nonstationary Model} \label{sec:methods-nonpooled-nonstationary}

To benchmark the extent to which spatial pooling improves the performance of our proposed model, we introduce the \textbf{Nonpooled Nonstationary Model}, in which we  conduct a process-informed nonstationary analysis for each station separately.

Previous studies in this region have used various climate predictors within the process-informed nonstationary frameworks.
These include thermodynamic covariates related to temperature changes such as global or regional temperature \citep{oldenborgh_attribution:2017, russell_gmxsst:2020}, dynamic changes related covariates like Ni\~{n}o~3.4 index \citep{risser_Harvey:2017}, and climate forcing factors, particularly $\coo$ concentrations \citep{risser_Harvey:2017, nielsen-gammon_harris:2020}.

In this study, we utilize global average $\coo$ concentration as the climate covariate, serving as a proxy for long-term global warming while minimizing noise from natural variability and volcanic eruptions \citep{jorgensen_nonstationary:2024}.
Since natural variability, like \gls{enso}, is related to tropical cyclone events, eliminating this factor helps us distinguish the impacts attributed solely to anthropogenic climate change warming.
We extract $\coo$ data from two sources:
monthly mean $\coo$ measured at Mauna Loa Observatory \citep{keeling_maunaloa:1976} dating back to 1958, and $\coo$ records derived from ice cores at Law Dome \citep{rubino_lawdome:2019} from \num{1006} to \num{1978}.
$\coo$ data from \num{1958} to \num{1978} are derived as the average of these two resources.

\begin{figure}[h]
    \centering
    \includegraphics[width=0.7\textwidth]{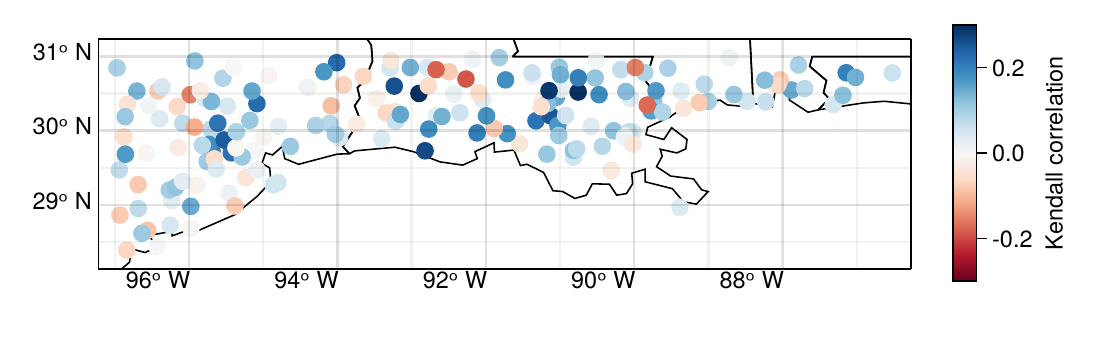}
    \caption{
        Annual maximum precipitation is positively correlated to $\ln\coo$ at most stations.
        Each dot represents a station analyzed.
        Color indicates the rank correlation (Kendall's $\tau$-b) between $\ln\coo$ and annual maximum precipitation at each station.
        Blue (red) indicates positive (negative) relationships.
    }\label{fig:co2-cor}
\end{figure}

Specifically, we assume the \gls{gev} location and scale parameters are conditioned on $\ln\coo$ (\Cref{eqn:nonpooled-nonstationary-location,eqn:nonpooled-nonstationary-scale}).
Allowing location and scale parameters to vary simultaneously enables the distributions to both expand and shift to account for observed nonstationarity \citep{nielsen-gammon_harris:2020}.
Because shape parameters are hard to estimate, even in a stationary case \citep{coles_extremes:2001}, here we neglect temporal nonstationarity of shape parameters, consistent with the practices from some other previous research \citep{ragno_processinformed:2019, song_extremechina:2020, jorgensen_nonstationary:2024}.
And we estimate the shape parameters at each station separately.
Our model is thus given by:
\begin{align}
    y(s,t)      & \sim \mathrm{GEV}(\mu(s,t), \sigma(s,t), \xi(s)) \label{eqn:nonpooled-nonstationary-model}        \\
    \mu(s,t)    & = \alpha_\mu (s) + \beta_\mu(s)x(t) \label{eqn:nonpooled-nonstationary-location}                  \\
    \sigma(s,t) & = \exp \{ \log\alpha_\sigma(s) + \beta_\sigma(s)x(t) \} \label{eqn:nonpooled-nonstationary-scale} \\
    \xi (s, t)  & = \xi \label{eqn:nonpooled-nonstationary-shape}
\end{align}
in which, $\alpha_\mu$ and $\alpha_\sigma$ are the regression intercepts, $\beta_\mu$ and $\beta_\sigma$ are the regression coefficients, and $x(t)$ is the time-varying climate covariate.

Within the Bayesian framework, we select priors based on general knowledge about the magnitude of extreme precipitation, avoiding strong priors that could overly bias the simulation results.
For instance, daily \gls{ams} typically ranges from \SIrange{2}{20}{inch}, so we assign a Normal prior with a mean of \num{5} and a standard deviation of \num{5} to the location intercept.
The prior for the shape parameter aligns with the stationary model (\Cref{eqn:GEV-shape-prior}).
Other remaining priors are set to be relatively uninformative.
This yields a prior layer of:
\begin{align}
    \alpha_\mu        & \sim \mathrm{Normal}(5,5) \label{eqn:nonpooled-nonstationary-location-prior}             \\
    \log\alpha_\sigma & \sim \mathrm{Normal}(0,1) \label{eqn:nonpooled-nonstationary-logscale-prior}             \\
    \beta_\mu         & \sim \mathrm{Normal}(0,1) \label{eqn:nonpooled-nonstationary-location-coefficient-prior} \\
    \beta_\sigma      & \sim \mathrm{Normal}(0,1) \label{eqn:nonpooled-nonstationary-scale-coefficient-prior}
\end{align}
Because estimation is conducted for each station separately, the prior is the same for each station considered, and there is no spatial layer.

\subsubsection{Spatially Varying Covariates Model} \label{sec:methods-spatially-varying-covariate}

The \textbf{Spatially Varying Covariates Model} extends on the benchmark models, combining spatial pooling and process-informed method to estimate robust daily extreme precipitation frequencies.
Our final model is built upon the primary assumption that climate indices' impacts on heavy rainfall probabilities are spatially coherent, with nearby locations exhibiting similar spatiotemporal characteristics.

Explicitly, \gls{gev} location and scale parameters vary in space and time, and are represented with spatially varying parameters and time-varying variables.
Similar to the benchmark models, shape parameter is assumed to be constant across time and space.
In \cref{eqn:nonpooled-nonstationary-location,eqn:nonpooled-nonstationary-scale}, we assume that the regression intercepts ($\mu_0$, $\log\sigma_0$) and coefficients ($\beta_\mu$, $\beta_\sigma$) vary smoothly in space and are described with the latent \gls{gp} layer(\cref{eqn:pooled-nonstationary-location_intercept,eqn:pooled-nonstationary-scale_intercept,eqn:pooled-nonstationary-location_intercept_coeff,eqn:pooled-nonstationary-scale_coeff}).

For the Bayesian priors, kernel variance ($\alpha_k$) and length ($\rho_k$) parameters for all the spatially varying parameters (in \cref{eqn:nonpooled-nonstationary-location,eqn:nonpooled-nonstationary-scale}, $\mu_0$, $\log\sigma_0$, $\beta_\mu$ and $\beta_\sigma$) follow \textit{Inverse Gamma} and \textit{Gamma} distributions (\cref{eqn:GP-kernel-variance-prior,eqn:GP-kernel-length-prior}).
We set the \gls{gp} mean ($m_k$) as \num{0} as the Pooled Stationary Model.
As a result, the parameter priors are set to match those used in the pooled stationary model and the nonpooled nonstationary model.

\begin{align}
    y(s,t)          & \sim \mathrm{GEV}(\mu(s,t), \sigma(s,t), \xi) \label{eqn:pooled-nonstationary-model}               \\
    \mu_0(s)        & \sim \mathrm{GP}(0, K_{\mu_0}(s,s')) \label{eqn:pooled-nonstationary-location_intercept}           \\
    \log\sigma_0(s) & \sim \mathrm{GP}(0, K_{\log\sigma_0}(s,s')) \label{eqn:pooled-nonstationary-scale_intercept}       \\
    \beta_\mu(s)    & \sim \mathrm{GP}(0, K_{\beta_\mu}(s,s')) \label{eqn:pooled-nonstationary-location_intercept_coeff} \\
    \beta_\sigma(s) & \sim \mathrm{GP}(0, K_{\beta_{\sigma_0}}(s,s')) \label{eqn:pooled-nonstationary-scale_coeff}       \\
    \alpha_k        & \sim \mathrm{InverseGamma}(5,5) \label{eqn:GP-kernel-variance-prior}                               \\
    \rho_k          & \sim \mathrm{Gamma}(5,1) \label{eqn:GP-kernel-length-prior}                                        \\
    \xi             & \sim \mathrm{Normal}(0,0.5) \label{eqn:GEV-shape-prior}.
\end{align}

\subsubsection{Experiment Design} \label{sec:methods-experiment-design}

We fit the three aforementioned frameworks to daily \gls{ams} data, evaluate their performance through cross-validation, and compare the heavy rainfall estimates with those from \gls{noaa} Atlas 14 \citep{perica_atlas14_texas:2018}.
\Cref{table:comparison} summarizes the similarities and differences between our three models with respect to key features.
We further compare these three models to \gls{noaa} Atlas 14 for reference \citep{perica_atlas14_texas:2018}.

\begin{table}[htbp]
    \centering
    \caption{Comparison of extreme precipitation analysis frameworks}
    \label{table:comparison}
    \renewcommand{\arraystretch}{1.2} % Adjust row spacing for readability
    \setlength{\tabcolsep}{4pt} % Reduce column spacing for compactness
    \begin{tabular}{l p{2.5cm} p{2.5cm} p{3cm} p{3cm}} 
        \hline
                      & \textbf{\gls{noaa} Atlas 14} & \textbf{Pooled Stationary} & \textbf{Nonpooled Nonstationary} & \textbf{Spatially Varying Covariate} \\ 
        \hline
        Stationarity  & Stationary model & Stationary model & Process-informed nonstationary & Process-informed nonstationary \\ 
        \hline
        Interpolation & Region of influence & Spatially pooled GEV & Each station separately & Spatially pooled climate impact parameters \\ 
        \hline
        Computation   & $L$-moments & Bayesian & Bayesian & Bayesian \\ 
        \hline
    \end{tabular}
\end{table}

\subsection{Validation} \label{sec:methods-validation}

We conduct cross-validation in terms of time and space to evaluate the model performance.

First, to compare the three different frameworks outlined in \cref{sec:proposed-framework}, we exclude observations from even (odd) years to assess the out-of-sample predictability on observations from odd (even) years.
All stations are utilized to keep consistency since the Nonpooled Nonstationary Model simulates each station separately.
To assess the out-of-sample predictive performance, we utilize the \gls{logs}, \gls{qs}, and \gls{crps} to validate the probabilistic projections.
For each validation metric, we first calculate the average across all simulated posterior distributions for a single observation, and then compute the average across all observation records.

Second, we conduct a spatial cross-validation of the Spatially Varying Covariates Model to assess how well the model is able to extrapolate to unobserved locations.
Specifically, to avoid including nearby stations in the same subset, the study region is first partitioned into \num{25} grids (\cref{fig:S1_Station_subsets}).
For each subset, five diagonal grids are excluded, and the model is fitted repeatedly to the remaining \num{20} grids of stations.
\Cref{fig:S2_raw_data_nobs_subs} shows the stations used to fit the model for the five subsets along with all available stations.
The estimates for the excluded stations are compared with those obtained using all stations.
This allows us to assess the model's ability to extrapolate not only to unobserved locations, but also to stations without any nearby observations.

We use several validation scores to assess the model performance.
\Gls{logs} \citep{selten_quadraticscoring:1998, gelman_bda3:2014} calculates the negative logarithm of the posterior predictive density.
It assesses the overall likelihood of the observations given the simulated distributions, and lower \gls{logs} value is often desired.
For an observation $y_i$ at station $s$ in year $t$ given all the posterior simulations, the \gls{logs} is:
\begin{equation}
    \label{eqn:equation14}
    -\log p_\mathrm{post}(y_i) = 
    -\log\mathbb{E}_\mathrm{post}(p(y_i|\theta)) = 
    -\log\int p(y_i|\theta)p_\mathrm{post}(\theta)\,\mathrm{d}\theta
\end{equation}
where $p_\mathrm{post}(y_i)$ is the predictive density for observation $y_i$ given the simulated posterior distributions, $\mathbb{E}_\mathrm{post}(p(y_i|\theta))$ is the expected value of the likelihood $p(y_i|\theta)$ with respect to the posterior distribution parameters $\theta$, $p_\mathrm{post}(\theta)$ is the posterior distribution, and $\int p(y_i|\theta)p_\mathrm{post}(\theta)\,\mathrm{d}\theta$ is the average of likelihood given all posterior distributions.

The \gls{qs} \citep{ulrich_idf:2020, bentzien_quantilescore:2014, koenker_goodnessfit:1999} measures the weighted difference between observation and the modeled quantiles.
It assesses how well the simulations match observations at different parts of the distribution, especially useful for extremes.
Typically, a smaller \gls{qs} value indicates a more reliable model.
The quantile score for an observation $y_i$ for a particular non-exceedance probability $p$ is calculated:
\begin{align}
    \label{eqn:equation15}
    \gls{qs}(p) & = \frac{1}{N}\sum_{n=1}^{N}\rho_p(y_i - q_{p,n}) \\
    \label{eqn:equation16}
    \rho_p(u)   & = \begin{cases}
                        pu     & \text{if } u > 0    \\
                        (p-1)u & \text{if } u \leq 0
                    \end{cases}
\end{align}
where $N$ is the total number of simulations, $q_{p,n}$ is the corresponding modeled quantiles at probability $p$, and $\rho_p$ is the loss function.

\Gls{crps} \citep{brocker_CRPS:2012} quantifies the integrated squared difference between the predicted \gls{cdf} and the actual observation.
It evaluates the overall accuracy of simulated distributions.
\gls{crps} close to \num{0} indicates better accuracy.
It is given by:
\begin{equation}
    \label{eqn:equation17}
    \mathrm{CRPS}(F,y_i) = \int_{-\infty}^{\infty} \qty[ F(x) - \mathbb{I}\{x \geq y_i\} ] ^2\, \dd{x}
\end{equation}
where $F$ is the \gls{cdf} of $X$, i.e. $F(x) = P[X \geq y]$; $y_i$ is the observation; $\mathbb{I}(x)$ is \num{1} if the real argument is positive or zero, \num{0} otherwise.

\subsection{Estimation} \label{sec:methods-estimation}

The field of hydrological frequency analysis has developed a wide range of methods for estimating model parameters.
Traditionally, point estimators including methods of moments, \gls{mle} \citep{martins_estimators:2000, stedinger_estimators:1997}, and $L$-moments have been widely used \citep{perica_atlas14_texas:2018, hosking_lmoments:1990, hosking_rfa:1997}.
However, moment-based methods have limited flexibility to be integrated with the process-informed framework \citep{cooley_spatial:2007, katz_hydroextrems:2002}.
In contrast, Bayesian statistical analysis offers a more adaptable approach, enabling the incorporation of nonstationary and regionalization through a hierarchical Bayesian framework \citep{ragno_processinformed:2019, thiemann_hydrobayes:2001}.
This method provides interpretable probabilistic predictions by updating prior knowledge of unknown quantities using observational data.
Specifically, the posterior probability is derived from the prior probability and the likelihood function \citep{gelman_bda3:2014}.
Additionally, Bayesian models can be effectively diagnosed with tools such as trace plots and Rhat values, ensuring model convergence and reliability \citep{gelman_workflow:2020}.

Given a model and data, we use \gls{mcmc} to estimate the posterior distribution of model parameters.
To speed up convergence \citep{cooley_spatial:2007}, we set the initial values as the \gls{map} results.
The Bayesian framework is built in R and the stan programming language,, and we employ the \gls{nuts} simulator \citep{stan_userguide:2022}.
A total of \num{4} chains each of \num{10000} samples are simulated, when the first half is set as the ``warm-up'' phase.
Posterior \gls{gev} parameters are used to compute return levels for all \gls{mcmc} simulations and are further analyzed to determine uncertainty ranges and expected estimates.

%% ------------------------------------------------------------------------ %%
%
%  RESULTS
%
%% ------------------------------------------------------------------------ %%

\section{Results} \label{sec:results}

We fit \gls{ams} of daily precipitation from the selected stations to each of the three models: Pooled Stationary Model, Nonpooled Nonstationary Model, and Spatially Varying Covariates Model.
In \cref{sec:results-performance} we assess the performance of the Spatially Varying Covariates Model.
In \cref{sec:results-patterns} we show the spatial pattern of estimated model parameters, return level, and trends over time at each station.
In \cref{sec:results-comparison} we compare the Spatially Varying Covariates Model to the two alternatives and with \gls{noaa} Atlas 14 estimates.

\subsection{Model performance} \label{sec:results-performance}

A critical question is whether our proposed model accurately and robustly captures nonstationary precipitation probabilities.
We assess the performance of the Spatially Varying Covariates Model using three approaches.
First, we use standard \gls{mcmc} diagnostics to assesss the convergence and sampling of our model \citep{gelman_workflow:2020}.
\Cref{fig:S3_traceplots} shows the trace plots of key distribution parameters, which appear to be qualitatively well-mixed.
Additionally, $\hat{R}$ values close to \num{1} are consistent with sufficient sampling and convergence \cite[chapter 11]{gelman_bda3:2014}.

\begin{figure}[h]
    \centering
    \includegraphics[width=0.7\textwidth]{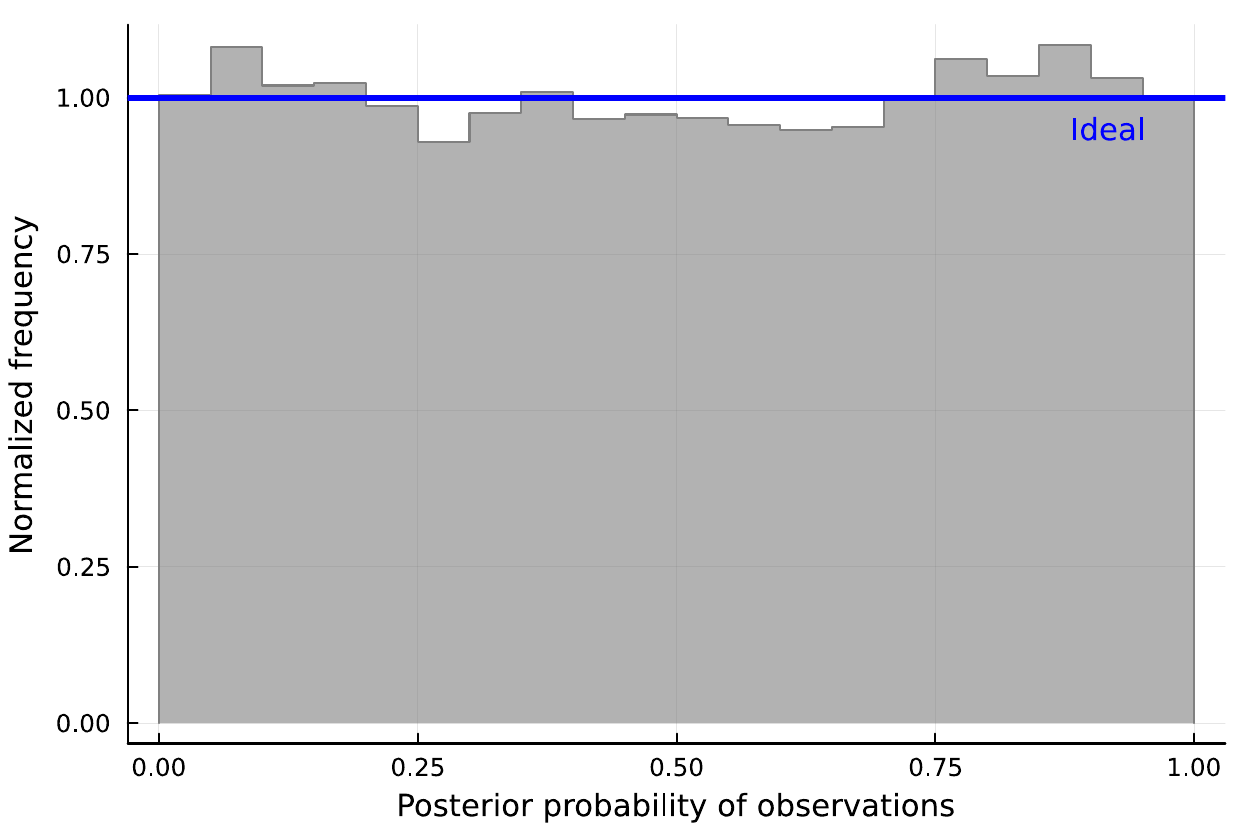}
    \caption{
        The observations are well-distributed across the quantiles of the posterior distributions, indicating that the Spatially Varying Covariates Model can effectively capture the variability in the observations.
        Histogram plot shows the posterior probability of observations given the posterior GEV distributions.
    }\label{fig:quantile_plot}
\end{figure}

Next, we develop a graphical diagnostic to assess the quality of our fit.
Tools like residual plots and quantile-quantile plots are widely used in regression and extreme value analysis to assess model fit.
However, these tools are poorly suited for diagnosing nonstationary extreme value models across multiple locations.
Our model is spatiotemporal, meaning that we estimate a unique \gls{gev} distribution for each station and year.
For each observation, we compute the posterior probability of the observation given this posterior distribution.
If our model is well-calibrated, samples should appear in the bottom tenth of their respective posterior distributions as frequently as they do in the top tenth.
In other words, the observations should be uniformly distributed across all quantiles of the posterior distributions.
\Cref{fig:quantile_plot} shows the histogram of the posterior probability of the observations given the posterior GEV distributions.
The histogram generally displays nearly uniform shape, suggesting that the  model is well-calibrated, both for the overall distribution and especially for the upper tail.

Third, we evaluate how well the Spatially Varying Covariates Model predicts out-of-sample heavy rainfall probabilities.
To do this, we use $K=5$-fold cross-validation.
First, we divide stations into five spatially structured subsets, as discussed in \cref{sec:methods-validation} and illustrated in \cref{fig:S2_raw_data_nobs_subs}.
Then, we fit the model five times, leaving out one of the subsets each time, and then compare the estimates obtained from these subsets with those using all station observations.

\begin{figure}[h]
    \centering
    \includegraphics[width=1\textwidth]{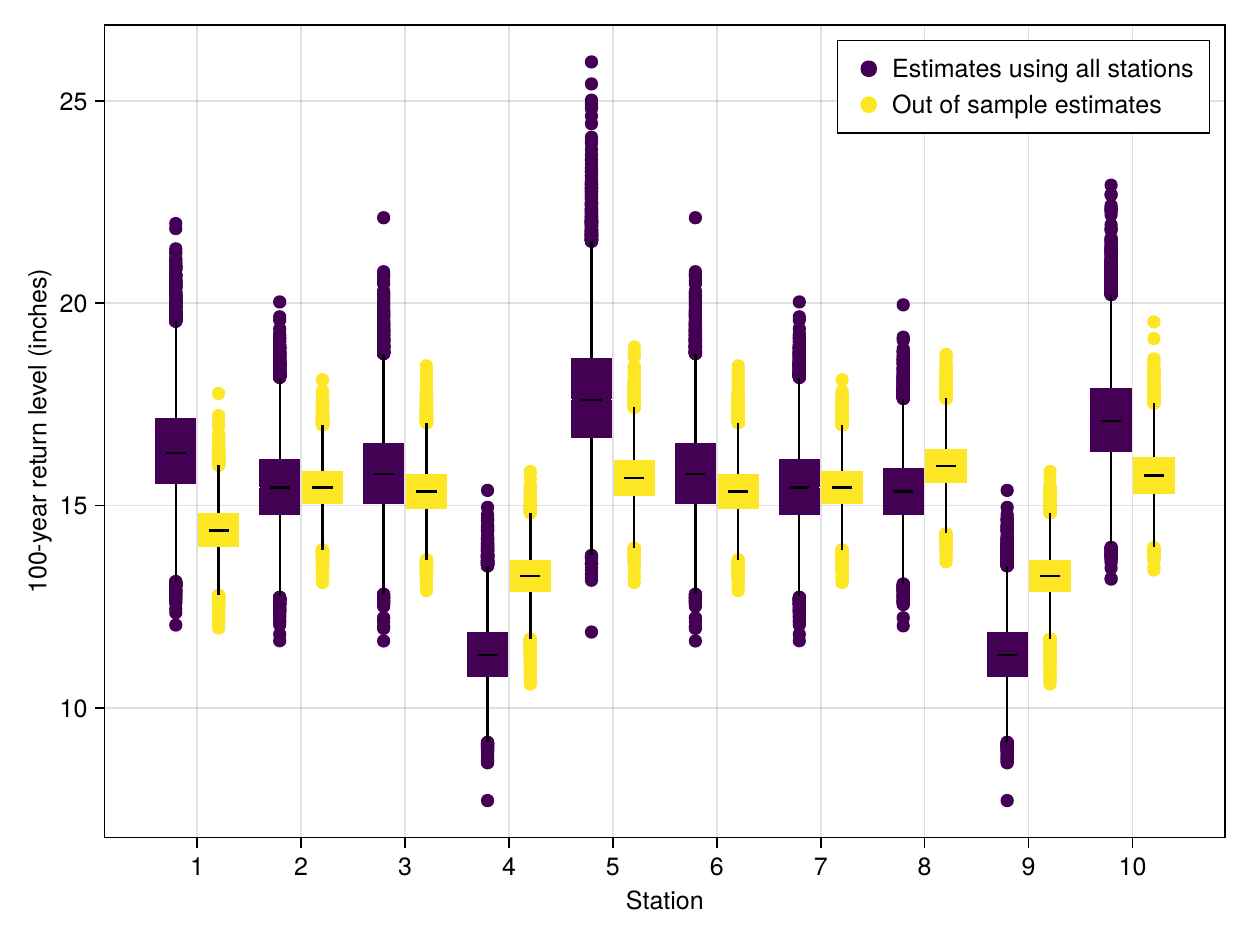}
    \caption{
        Out-of-sample probabilistic inferences generally agree with those from the full dataset.
        The plot shows posterior estimates of \num{100}-year return level in \num{2022} when (dark) using all data (light) and using station subsets (out-of-sample).
        We choose to present results for \num{2} stations from each of the \num{5} subsets, representing different areas of the study region.
    }\label{fig:outS-boxplots}
\end{figure}

\begin{figure}[h]
    \centering
    \includegraphics[width=1\textwidth]{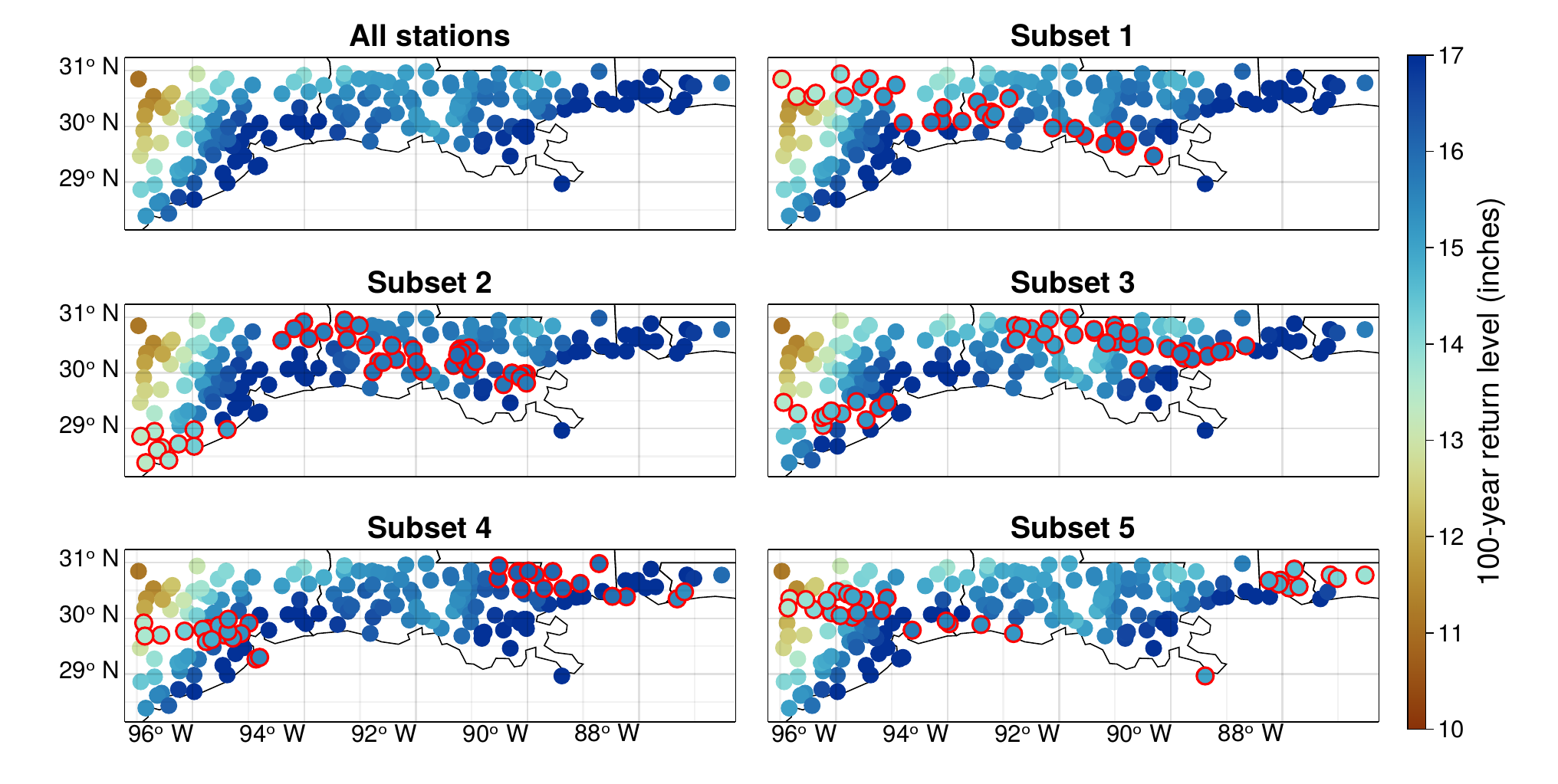}
    \caption{
        Out-of-sample estimates generally match estimates with full dataset.
        The maps present \num{100}-year return level estimates at all stations when using all stations and station subsets.
        The circled stations are excluded for the specific station subset (out-of-sample).
    }\label{fig:outS-maps}
\end{figure}

\Cref{fig:outS-boxplots} compares the posterior distribution of \num{100}-year return level when using the full model and out-of-sample estimates for ten representative stations (two from each of the five subsets).
Although some differences between the posterior distributions are evident, as expected, the posterior means are very similar for most stations.
\Cref{fig:outS-maps} presents the 100-year return level estimates (for \num{2022} $\coo$ levels) for all stations, comparing simulations using all stations to those from the \num{5} subsets.
Both the spatial patterns and the magnitude of return levels are generally consistent across the six (five subsets plus the full model) simulations.
These findings support to the model's ability to robustly predict nonstationary return levels at unobserved locations

\subsection{Spatial and temporal patterns of daily extreme precipitation} \label{sec:results-patterns}

A second key question is how the probabilities of extreme precipitation vary in space and time.

\subsubsection{Robust nonstationary parameter estimates} \label{sec:results-CO2-coef}

\begin{figure}[h]
    \centering
    \includegraphics[width=1\textwidth]{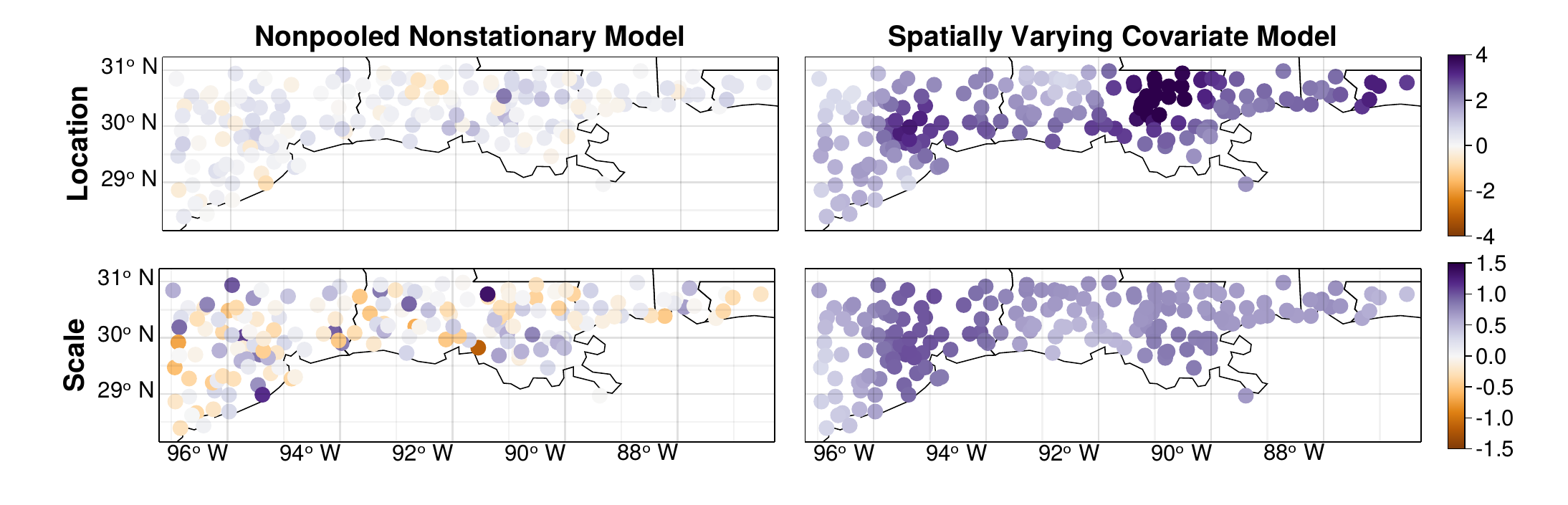}
    \caption{
        Climate change contributes to increases in intensity and variability of extreme precipitation. Figure shows the posterior mean of the coefficients of $\ln\coo$ on (T) GEV location ($\beta_\mu$ in \Cref{eqn:nonpooled-nonstationary-location}) and (B) \gls{gev} scale ($\beta_\sigma$ in \Cref{eqn:nonpooled-nonstationary-scale}) parameters based on the (L) Nonpooled Nonstationary Model and (R) Spatially Varying Covariates Model.
    }\label{fig:co2-coef}
\end{figure}

We employ the process-informed nonstationary framework, wherein \gls{gev} location and scale parameters are treated as functions of global mean $\coo$.
However, as discussed in \cref{sec:introduction}, a longstanding critique of nonstationary models is practical in nature, relating to the large estimation uncertainty associated with estimating more parameters from the same data \citep{serinaldi_undead:2015}.
To assess whether the Spatially Varying Covariates Model can robustly estimate the nonstationary parameters, we compare the coefficients estimated from the Spatially Varying Covariates Model and the Nonpooled Nonstationary Model.
As shown in \cref{fig:co2-coef}, the Nonpooled Nonstationary Model yields estimates that vary substantially even between nearby stations, similar to previous efforts \citep{fagnant_spatiotemporal:2020}.
In contrast, the Spatially Varying Covariates Model yields spatially smooth coefficient estimates.
Across the majority of the study area, $\ln\coo$ demonstrates a positive correlation with both GEV parameters.
This indicates that as global $\coo$ concentration rise, both the \gls{gev} location and scale parameters increase, indicating an increase in the overall intensity and variability in the extreme precipitation.
The highest coefficient estimates are found around Houston and New Orleans, implying that these regions have the strongest increasing trends in extreme precipitation.
It is worth noting that if we use a semi-Bayesian framework, where point estimates are developed separately at each station and then smoothed as in \cite{ossandon_semibayesian:2021}, the coefficient estimates would differ significantly from those of our fully spatial model.

\subsubsection{Changes in return levels} \label{sec:spatio-temporal-patterns}

\begin{figure}[h]
    \centering
    \includegraphics[width=1\textwidth]{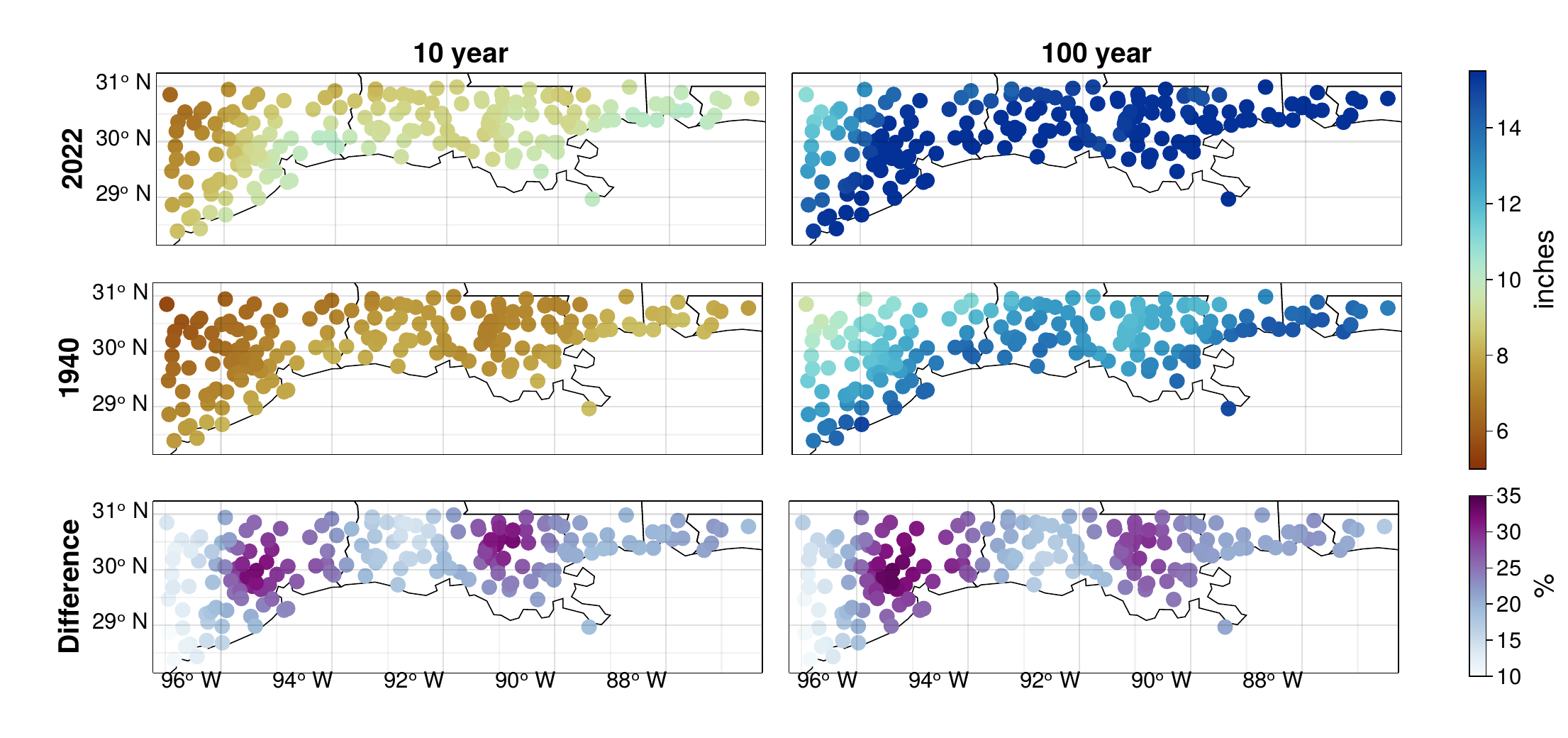}
    \caption{
        The Spatially Varying Covariates Model projects spatially consistent increases in daily heavy rainfall probabilities in the western Gulf Coast.
        These maps show the posterior mean of the (L) \num{10}-year and (R) \num{100}-year extreme precipitation return level estimates (T) in \num{2022} (M) in \num{1940} (B) percentage change from \num{1940} to \num{2022}.
    }\label{fig:spatio-temporal-patterns}
\end{figure}

\Cref{fig:co2-coef} indicates that both the location and scale parameters exhibit sensitivity to global $\coo$ concentration.
However, applications of precipitation extremes are often based on return levels, which are the expected value of the precipitation amount that is exceeded with a given probability.
To analyze the spatiotemporal characteristics of extreme precipitation probabilities, we plot the return level estimates in \num{2022} and \num{1940}, as well as the percentage change from \num{1940} to \num{2022} in \cref{fig:spatio-temporal-patterns}. 
We also generate gridded estimates by interpolating distribution parameters to grid center (\cref{sec:background-GP}), shown in \cref{fig:S4_gridded}.
We find that return levels have increased by between \num{10} and \num{35}$\%$ over the past \num{80} years throughout the study region, with the largest increases observed in coastal Southeast Texas and coastal southeastern Louisiana.

\subsection{Model comparison} \label{sec:results-comparison}

Another key question is how robust the Spatially Varying Covariates Model is compared to other frameworks that do not incorporate both nonstationarity and regionalization, and how the nonstationary extreme rainfall probability estimates compare to stationary estimates, such as those in \gls{noaa} Atlas 14.

\subsubsection{Out-of-sample validations} \label{sec:OS-validations}
To assess the out-of-sample predictability of the three frameworks (\cref{sec:proposed-framework}), we carry out cross-validation by analyzing two subsets of the full data.
Specifically, we model with odd (even) year observation records and validate the projections for the excluded even (odd) years with metrics outlined in \cref{sec:methods-validation}.
Validation results are shown in \cref{table:validation}.
Overall, the Spatially Varying Covariates Model performs similarly to the Pooled Stationary Model and is notably better than the Nonpooled Nonstationary Model.

\gls{logs} evaluates the probability of the observations based on the simulated posterior distributions.
Since it measures negative log-likelihood, a lower \gls{logs} is desired, which indicates higher probability.
The Pooled Stationary Model achieves the lowest \gls{logs}, suggesting that its predicted probabilities are most consistent with the observed \gls{ams}, and therefore provides a better match between the simulations and the actual observations.
The Spatially Varying Covariates Model performs similarly well.
However, the Nonpooled Nonstationary Model exhibits a significantly higher \gls{logs}.
This model analyzes each station separately, and when half of the years are excluded, some stations are left with insufficient data (as few as \num{15} years of observations).
This leads to high uncertainty in the estimates.
Additionally, \gls{logs} heavily penalizes the unlikely events \citep{bjerregard_validation:2021}.
As a result, the Nonpooled Nonstationary Model may generate distributions that predict some observations as extremely rare events, significantly increasing the \gls{logs} value.

\gls{qs} measures how well the predicted quantiles match the observed values, and a smaller \gls{qs} indicates better performance.
Specifically, we evaluate extreme rainfall estimates with non-exceedance probabilities of \num{0.9}, \num{0.98}, \num{0.99} (corresponding to \num{10}-, \num{50}- and \num{100}-year events).
The two spatially pooled models show similar performance, with the Pooled Stationary Model achieving the lowest \gls{qs} value at the non-exceedance probability of \num{0.9}.
The Spatially Varying Covariates Model, on the other hand, performs best at higher non-exceedance probabilities (\num{0.98} and \num{0.99}), indicating that this model better captures more extreme events.
The Nonpooled Nonstationary Model yields the highest \gls{qs} values, mainly due to the high uncerainties in the estimated quantiles.

\gls{crps} provides a comprehensive assessment of simulation accuracy across all probabilities of the posterior distributions, with values close to \num{0} being preferred.
In general, this metric favors models when observations are near the median of the simulated distributions.
Similar to \gls{logs}, the Pooled Stationary Model and the Spatially Varying Coviates Model show comparable \gls{crps} values, though the former achieves slightly smaller values.
These three metrics collectively indicate that the Spatially Varying Covariates Model outperforms the nonstationary framework at individual stations, while being comparable to the stationary framework.
This suggests that integrating both nonstationary and regionalization could enhance nonstationary extreme precipitation analysis.

\begin{table}[htbp]
    \centering
    \caption{Comparison of extreme precipitation analysis frameworks}
    \label{table:validation}
    \renewcommand{\arraystretch}{1.2} % Adjust row height for better readability
    \setlength{\tabcolsep}{6pt} % Reduce column spacing for compactness
    \begin{tabular}{l c c c} 
        \hline
                      & \textbf{Pooled Stationary} & \textbf{Nonpooled Nonstationary} & \textbf{Spatially Varying Covariate} \\ 
        \hline
        LogS          & 1.9322 & 1.9903 & 1.9495 \\ 
        QS ($p = 0.9$)  & 0.4671 & 0.5211 & 0.4712 \\ 
        QS ($p = 0.98$) & 0.1689 & 0.2219 & 0.1680 \\ 
        QS ($p = 0.99$) & 0.1027 & 0.1508 & 0.1018 \\ 
        CRPS          & 0.2548 & 0.2809 & 0.2574 \\ 
        \hline
    \end{tabular}
\end{table}

\subsubsection{Comparison with \acrshort{noaa} Atlas 14 estimates} \label{sec:Atlas14-comparison}
\begin{figure}[h]
    \centering
    \includegraphics[width=1\textwidth]{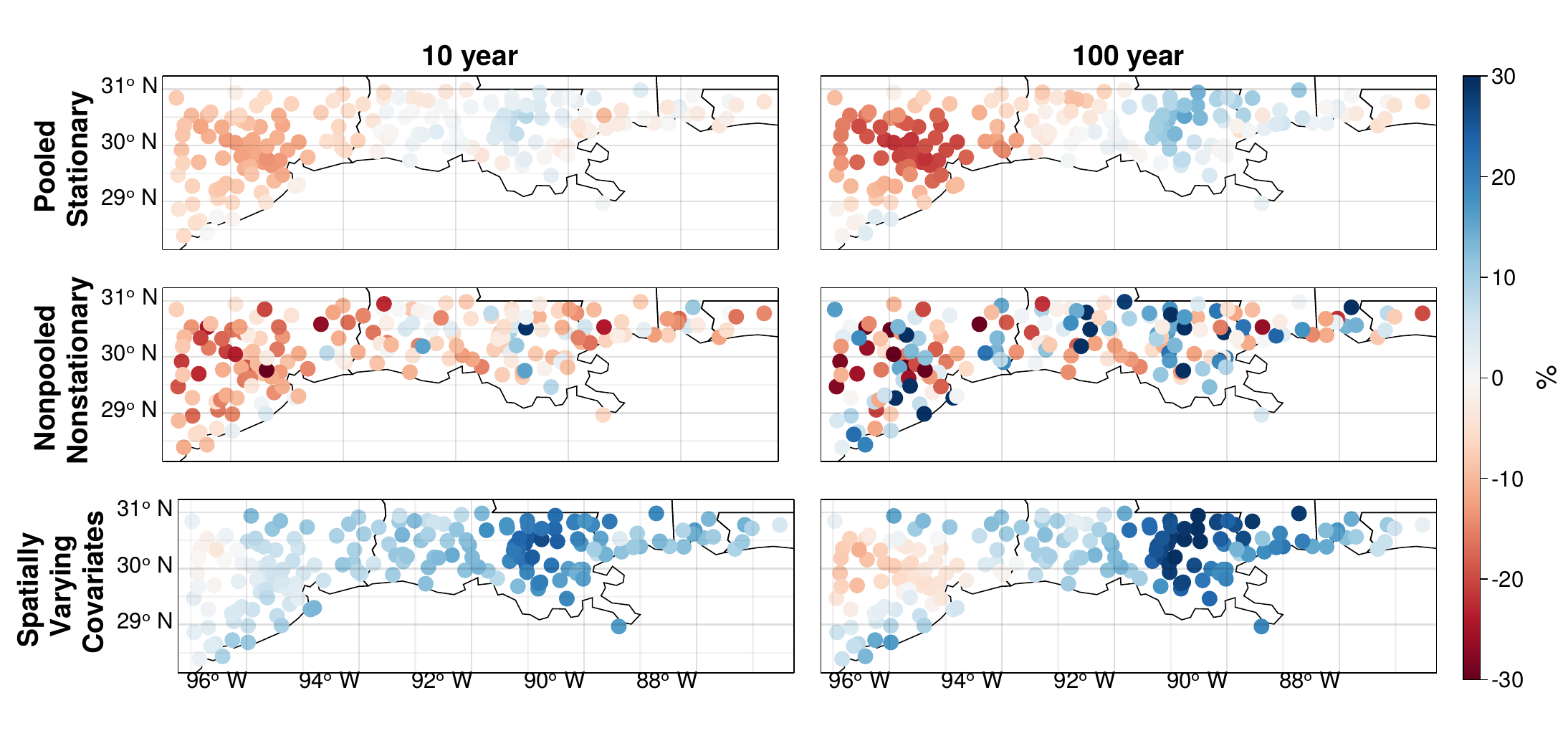}
    \caption{
        The Spatially Varying Covariates Model yields higher estimates in the eastern part of the study area and lower estimates in the western part compared to \gls{noaa} Atlas 14 estimates.
        The percentage difference between the estimates from \gls{noaa} Atlas 14 and our \num{2022} model implementation is illustrated as follows: (T) the Pooled Stationary Model, (M) Nonpooled Nonstationary Model, and (B) Spatially Varying Covariates Model for (L) \num{10}-year and (R) \num{100}-year events.
        Positive (blue) values indicate our estimates are higher, and negative (red) values indicate that our estimates are lower than \gls{noaa} Atlas 14 estimates.
        The percentage difference is calculated as: (our estimates in \num{2022} - \gls{noaa} Atlas 14 estimates)/\gls{noaa} Atlas 14 estimates * \num{100}$\%$.}
    \label{fig:Atlas14_diff}
\end{figure}

Many practitioners will be interested in differences between our estimates and those from the current guidance.
\Gls{noaa} Atlas 14 utilizes historical observations until \num{2017} and adopts a stationary framework with regionalization.
\Cref{fig:Atlas14_diff} presents the percentage difference between the posterior mean return level estimates in \num{2022} from the three models and those from \gls{noaa} Atlas 14.
In general, across these three frameworks, our estimates are higher than \gls{noaa} Atlas 14 estimates in the western part of the study area and lower in the eastern part.
This spatial pattern in percentage differences is particularly pronounced for the Pooled Stationary Model and more extreme events, as shown for the \num{100}-year events.
Although both are stationary frameworks, differences in the regionalization method and data length contribute to this variation.
The Nonpooled Nonstationary Model predominantly produces lower estimates for \num{10}-year events but higher estimates for \num{100}-year events.
This indicates that nonstationarity could contribute to more extreme events.
The Spatially Varying Covariates Model also tends to produce similar or lower estimates in the west, and higher estimates in the east, notably around western LA for \num{100}-year events.
The lower estimates around the Harris County region are consistent with some previous research \citep{nielsen-gammon_harris:2020, jorgensen_nonstationary:2024}.
\Gls{noaa} Atlas 14 utilizes observation data up to \num{2017}, which is \num{5} years shorter than our analysis, and includes the occurrence of Hurricane Harvey in \num{2017}.
Stations with short observation records may overfit to the \num{2017} events, potentially resulting in events like Harvey being estimated as an event with a shorter return period.

\begin{figure}[h]
    \centering
    \includegraphics[width=1\textwidth]{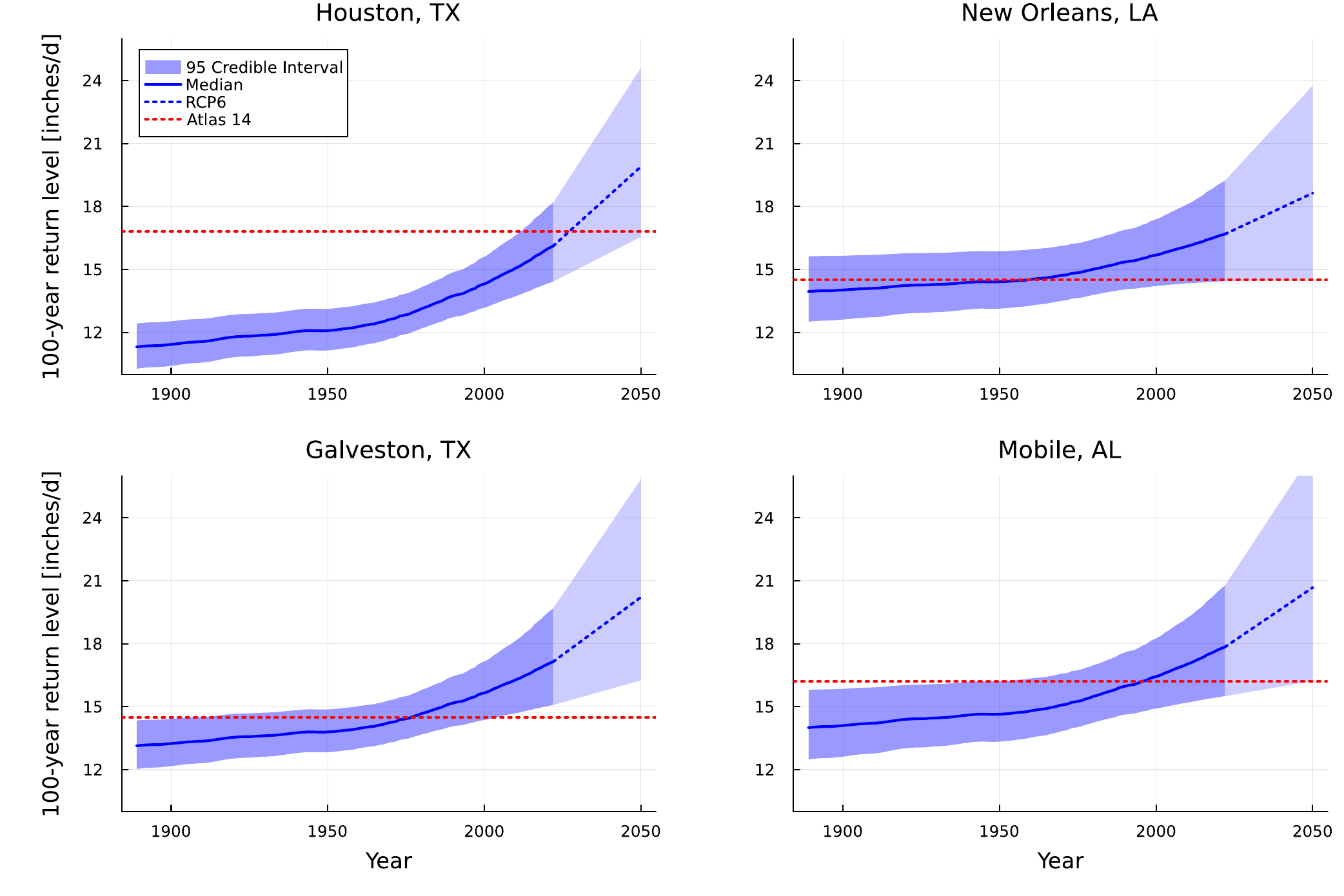}
    \caption{
        Future projections are higher than \gls{noaa} Atlas 14 estimates at major cities.
        Plots show the time series of \num{100}-year return level for daily precipitation in selected cities estimated from the Spatially Varying Covariates Model compared to Atlas 14 guidelines.
        Dashed blue lines use $\coo$ projections roughly corresponding to \gls{rcp} 6.
        Blue shadows are the \num{95}$\%$ credible interval from posterior estimates.
        }
    \label{fig:city-trends}
\end{figure}

We find that \gls{noaa} Atlas 14 underestimates return levels for \num{24}-hour precipitation in New Orleans, Galveston, and Mobile under current conditions and overestimates return levels for \num{24}-hour precipitation in Houston.
However, the lower estimates in Houston do not imply that \gls{noaa} Atlas 14 is adequate for engineering design in the face of future climate change.
\Cref{fig:city-trends} depicts the time series of return level estimates from the Spatially Varying Covariates Model, projecting to \num{2050} using $\coo$ concentrations from the \gls{rcp} \num{6} scenario.
We project that under this scenario, \gls{noaa} Atlas 14 will underestimate return levels in Houston after around \num{2025}.

%-----------------------------------------------------------------------------
% DISCUSSION -----------------------------------------------------------------
%-----------------------------------------------------------------------------
\section{Discussion} \label{sec:discussions}

We contribute to a growing and multidisciplinary literature on links between extreme precipitation and climate.
Methodologically, our reliance on observations to estimate nonstationary extreme preciptation draws primarily from engineering practice where accurate and local inferences are critical.
Downscaled and bias-corrected rainfall data from \glspl{gcm} offer broader spatial coverage and longer records, mitigating sampling variability in extreme rainfall probability estimates.
However, inherent dynamical biases in \glspl{gcm} limit their estimation reliability.
Our findings of increasing extreme precipitation probabilities across the Gulf Coast from historical observations are consistent with previous studies \citep{fagnant_spatiotemporal:2020, jorgensen_nonstationary:2024, statkewicz_raintx:2021}.

From a specific methodological perspective, the proposed Spatially Varying Covariates Model presents practical and theoretical advantages relative to other models for estimating nonstationary extreme precipitation probabilities from observations.
\cite{cheng_nonstationary:2014} examines nonstationary extreme rainfall probabilities at individual stations, highlighting the importance of incorporating climate change into design considerations.
\cite{ossandon_semibayesian:2021} builds a process-informed \gls{gev}, estimating in a two-step framework: first, point estimates of the distribution parameters are developed for each station separately, and then these noisy estimates are smoothed and interpolated with a \gls{gp}.
While this approach improves scalability by enabling independent and parallel estimation at each station, it introduces additional bias and uncertainty by separately interpolating each parameter, disregarding the correlations between them.
Alternatively, \cite{dyrrdal_bayesrain:2015} develops a Bayesian hierarchical model, when the \gls{gev} parameters are regressed on spatial and temporal covariates.
In this model, spatial coherence is achieved through the smoothness of the underlying spatial covariates.
An advantage is that it may be easier to account for topographic features that cause stations that are nearby in longitude and latitude to have different \gls{gev} parameters.
However, the choice of spatial covariates is challenging, as it's not immediately clear what the spatial covariates would be for a region like the Gulf Coast.
Finally, \cite{ulrich_idf:2020} employs a duration-dependent \gls{gev} distribution, modeling its parameters using polynomial basis functions within a stationary framework.

Several specific limitations of our model motivate further improvement and refinement.
First, alternative distributions besides \gls{gev} distribution can be explored, such as the blended \gls{gev} \citep{castro-camilo_bgev:2022}, which incorporates a \gls{gev} distribution with a \num{0} shape left tail and a positive shape right tail or the metastatistical distribution, which expands analysis beyond annual maxima \citep{miniussi_metastatistical:2020}.
Second, exploring how \gls{gp} kernels beyond the exponential capture diverse data characteristics and improve the simulations.
Third, improving estimation of parametric uncertainties.
In our model, the observed annual maxima at two nearby locations is considered independent, conditional on the estimated parameters.
However in reality, extreme events at adjacent locations are likely to be correlated.
Max-stable process models seek to address this limitation, but are computationally demanding and difficult to implement \citep{stephenson_maxstable:2016}.
Fourth, in regions with significant variations in topographic features, the inclusion of spatial variables such as elevation \citep{chu_elevation:2012, goovaerts_addelevation:2000, hu_elev:2021, talchabhadel_spatiotemporal:2018, dyrrdal_bayesrain:2015} may enhance model performance.
Fifth, one could extend our \gls{gp} layer to account for the dependency among \gls{gev} parameters through a coregionalization approach \citep{gelfand_spatial:2005, schmidt_coregionalization:2003}.
Sixth, one could model uncertainty in observed precipitation or incorporate data from alternative sources such as remote sensing or radar data \citep{gavahi_precipfusion:2023}.
Finally, because our model requires estimating parameters at each station simultaneously, and because \glspl{gp} estimation scales with the cube of the number of points, the computational complexity of our model may prove excessives for studies with many locations.
In such cases, alternative spatial pooling methods like basis functions \citep{ramsay_dataanalysis:2005} may offer computational efficiency while maintaining predictive quality.

\section{Conclusions} \label{sec:conclusions}
This study proposes a Bayesian hierarchical framework that integrates regionalization and nonstationarity to robustly estimate extreme precipitation probabilities.
Using a case study that models annual maximum precipitation on the Western Gulf Coast with a \gls{gev} distribution, the Spatially Varying Covariates Model (1) introduces process-informed nonstationarity, meaning that parameters are conditioned on climate covariates, and (2) assumes that parameters vary smoothly across space.
This framework naturally allows the use of gauges with incomplete observation records, pools information from nearby locations to generate smooth return level estimates, and can be applied to other regions, provided that suitable climate covariates are identified.

We demonstrate that the Spatially Varying Covariates Model provides robust estimates of return levels and their temporal trends.
Through rigorous cross-validation, we show that our estimates are both well-calibrated and reliable.
Specifically, the full model performs similarly to the stationary framework while outperforming the nonstationary framework at individual stations.
Additionally, it generates reliable estimates at ungauged locations. 

We find increase in \num{24}-hour extreme rainfall across the study area, particularly in the areas surrounding Houston and New Orleans.
Compared to the current guidance, Atlas 14, our estimates are generally lower in the western part and higher in the eastern part of the region.
Future projections for \num{2050}, based on \glspl{gcm}, suggest that return levels will exceed Atlas 14 estimates at most locations, underscoring the inadequacy of current guidelines for engineering design under a changing climate.

Nonstationarity is widely recognized, but the statistical challenges of robustly estimating nonstationary extreme precipitation probabilities have largely motivated continued use of stationary models.
This framework provides a practical and theoretically sound approach to estimating nonstationary extreme precipitation probabilities from observations.

%%% End of body of article

\section*{Data Availability Statement}

All data used in this research are publicly available. 
The daily rainfall data used to extract \gls{ams} records can be accessed from the \gls{ghcn} at \url{https://www.ncei.noaa.gov/products/land-based-station/global-historical-climatology-network-daily}. 
The $\coo$ data are sourced from Mauna Loa Observatory \citep{keeling_maunaloa:1976} and Law Dome \citep{rubino_lawdome:2019}, which are available at \url{https://gml.noaa.gov/ccgg/trends/data.html} and \url{https://data.csiro.au/collection/csiro%3A37077v3}, respectively.
Code to implement the Spatially Varying Covariates Model and the other frameworks, as well as plotting results can be accessed at the following github repository: \url{https://github.com/yuchenluv/SpatiallyVaryingCovariateModel}.

\section*{Acknowledgments}
This research was supported by Rice University and the Texas Water Development Board. 
The authors would like to thank Sylvia Dee, John Nielsen-Gammon, William Baule, Rewati Niraula, Saul Nuccitelli, and Srikanth Koka for insightful comments.

\bibliography{2024-yl-gulf-coast}

\begin{thebibliography}{101}
\providecommand{\natexlab}[1]{#1}
\providecommand{\url}[1]{\texttt{#1}}
\expandafter\ifx\csname urlstyle\endcsname\relax
  \providecommand{\doi}[1]{doi: #1}\else
  \providecommand{\doi}{doi: \begingroup \urlstyle{rm}\Url}\fi

\bibitem[Aich et~al.(2024)Aich, Hess, Pan, Bathiany, Huang, and Boers]{aich_landuse:2016}
Michael Aich, Philipp Hess, Baoxiang Pan, Sebastian Bathiany, Yu~Huang, and Niklas Boers.
\newblock Conditional diffusion models for downscaling \& bias correction of {{Earth}} system model precipitation, April 2024.

\bibitem[Ansh~Srivastava and Mascaro(2023)]{anshsrivastava_raedaridf:2023}
Nehal Ansh~Srivastava and Giuseppe Mascaro.
\newblock Improving the utility of weather radar for the spatial frequency analysis of extreme precipitation.
\newblock \emph{Journal of Hydrology}, 624:\penalty0 129902, September 2023.
\newblock ISSN 0022-1694.
\newblock \doi{10.1016/j.jhydrol.2023.129902}.

\bibitem[Apputhurai and Stephenson(2013)]{apputhurai_spatiotemporal:2013}
Pragalathan Apputhurai and Alec~G. Stephenson.
\newblock Spatiotemporal hierarchical modelling of extreme precipitation in {{Western Australia}} using anisotropic {{Gaussian}} random fields.
\newblock \emph{Environmental and Ecological Statistics}, 20\penalty0 (4):\penalty0 667--677, December 2013.
\newblock ISSN 1573-3009.
\newblock \doi{10.1007/s10651-013-0240-9}.

\bibitem[Bass et~al.(2017)Bass, Juan, Gori, Fang, and Philip]{bass_memorial:2017}
Benjamin Bass, Andrew Juan, Avantika Gori, Zheng Fang, and Bedient Philip.
\newblock 2015 {{Memorial Day Flood Impacts}} for {{Changing Watershed Conditions}} in {{Houston}}.
\newblock \emph{Natural Hazards Review}, 18\penalty0 (3):\penalty0 05016007, August 2017.
\newblock \doi{10.1061/(asce)nh.1527-6996.0000241}.

\bibitem[Bentzien and Friederichs(2014)]{bentzien_quantilescore:2014}
Sabrina Bentzien and Petra Friederichs.
\newblock Decomposition and graphical portrayal of the quantile score: {{Quantile Score Decomposition}} and {{Portrayal}}.
\newblock \emph{Quarterly Journal of the Royal Meteorological Society}, 140\penalty0 (683):\penalty0 1924--1934, July 2014.
\newblock ISSN 00359009.
\newblock \doi{10.1002/qj.2284}.

\bibitem[Bjerreg{\aa}rd et~al.(2021)Bjerreg{\aa}rd, M{\o}ller, and Madsen]{bjerregard_validation:2021}
Mathias~Blicher Bjerreg{\aa}rd, Jan~Kloppenborg M{\o}ller, and Henrik Madsen.
\newblock An introduction to multivariate probabilistic forecast evaluation.
\newblock \emph{Energy and AI}, 4:\penalty0 100058, June 2021.
\newblock ISSN 2666-5468.
\newblock \doi{10.1016/j.egyai.2021.100058}.

\bibitem[Blanchet et~al.(2016)Blanchet, Ceresetti, Molini{\'e}, and Creutin]{blanchet_invariant:2016}
J.~Blanchet, D.~Ceresetti, G.~Molini{\'e}, and J.-D. Creutin.
\newblock A regional {{GEV}} scale-invariant framework for {{Intensity}}--{{Duration}}--{{Frequency}} analysis.
\newblock \emph{Journal of Hydrology}, 540:\penalty0 82--95, September 2016.
\newblock ISSN 00221694.
\newblock \doi{10.1016/j.jhydrol.2016.06.007}.

\bibitem[Br{\"o}cker(2012)]{brocker_CRPS:2012}
Jochen Br{\"o}cker.
\newblock Evaluating raw ensembles with the continuous ranked probability score.
\newblock \emph{Quarterly Journal of the Royal Meteorological Society}, 138\penalty0 (667):\penalty0 1611--1617, July 2012.
\newblock ISSN 0035-9009, 1477-870X.
\newblock \doi{10.1002/qj.1891}.

\bibitem[Burn(1990)]{burn_regionofinfluence:1990}
Donald~H. Burn.
\newblock Evaluation of regional flood frequency analysis with a region of influence approach.
\newblock \emph{Water Resources Research}, 26\penalty0 (10):\penalty0 2257--2265, 1990.
\newblock ISSN 1944-7973.
\newblock \doi{10.1029/WR026i010p02257}.

\bibitem[Cannon et~al.(2015)Cannon, Sobie, and Murdock]{cannon_GCMbc:2015}
Alex~J. Cannon, Stephen~R. Sobie, and Trevor~Q. Murdock.
\newblock Bias {{Correction}} of {{GCM Precipitation}} by {{Quantile Mapping}}: {{How Well Do Methods Preserve Changes}} in {{Quantiles}} and {{Extremes}}?
\newblock September 2015.
\newblock \doi{10.1175/JCLI-D-14-00754.1}.

\bibitem[Carpenter et~al.(2017)Carpenter, Gelman, Hoffman, Lee, Goodrich, Betancourt, Brubaker, Guo, Li, and Riddell]{carpenter_stan:2017}
Bob Carpenter, Andrew Gelman, Matthew~D Hoffman, Daniel Lee, Ben Goodrich, Michael Betancourt, Michael~A Brubaker, Jiqiang Guo, Peter Li, and Allen Riddell.
\newblock Stan: A probabilistic programming language.
\newblock \emph{Journal Of Statistical Software}, 76\penalty0 (1):\penalty0 1--29, January 2017.
\newblock \doi{10.18637/jss.v076.i01}.

\bibitem[Castro-Camilo et~al.(2022)Castro-Camilo, Huser, and Rue]{castro-camilo_bgev:2022}
Daniela Castro-Camilo, Rapha{\"e}l Huser, and H{\aa}vard Rue.
\newblock Practical strategies for generalized extreme value-based regression models for extremes.
\newblock \emph{Environmetrics}, 33\penalty0 (6):\penalty0 e2742, September 2022.
\newblock ISSN 1180-4009, 1099-095X.
\newblock \doi{10.1002/env.2742}.

\bibitem[Cheng and AghaKouchak(2014)]{cheng_nonstationary:2014}
Linyin Cheng and Amir AghaKouchak.
\newblock Nonstationary precipitation intensity-duration-frequency curves for infrastructure design in a changing climate.
\newblock \emph{Scientific Reports}, 4\penalty0 (1):\penalty0 7093, November 2014.
\newblock ISSN 2045-2322.
\newblock \doi{10.1038/srep07093}.

\bibitem[Chu(2012)]{chu_elevation:2012}
Hone-Jay Chu.
\newblock Assessing the relationships between elevation and extreme precipitation with various durations in southern {{Taiwan}} using spatial regression models.
\newblock \emph{Hydrological Processes}, 26\penalty0 (21):\penalty0 3174--3181, October 2012.
\newblock ISSN 0885-6087, 1099-1085.
\newblock \doi{10.1002/hyp.8403}.

\bibitem[Coles(2001)]{coles_extremes:2001}
Stuart Coles.
\newblock \emph{An Introduction to Statistical Modeling of Extreme Values}.
\newblock Springer Series in Statistics. Springer, London ;, 2001.
\newblock ISBN 1-85233-459-2.

\bibitem[Cook et~al.(2020)Cook, McGinnis, and Samaras]{cook_stormwater:2020}
Lauren~M. Cook, Seth McGinnis, and Constantine Samaras.
\newblock The effect of modeling choices on updating intensity-duration-frequency curves and stormwater infrastructure designs for climate change.
\newblock \emph{Climatic Change}, 159\penalty0 (2):\penalty0 289--308, March 2020.
\newblock ISSN 1573-1480.
\newblock \doi{10.1007/s10584-019-02649-6}.

\bibitem[Cooley and Sain(2010)]{cooley_spatialhierarchical:2010}
Daniel Cooley and Stephan~R. Sain.
\newblock Spatial hierarchical modeling of precipitation extremes from a regional climate model.
\newblock \emph{Journal of Agricultural, Biological, and Environmental Statistics}, 15\penalty0 (3):\penalty0 381--402, September 2010.
\newblock ISSN 1537-2693.
\newblock \doi{10.1007/s13253-010-0023-9}.

\bibitem[Cooley et~al.(2007)Cooley, Nychka, and Naveau]{cooley_spatial:2007}
Daniel Cooley, Douglas Nychka, and Philippe Naveau.
\newblock Bayesian spatial modeling of extreme precipitation return levels.
\newblock \emph{Journal of the American Statistical Association}, 102\penalty0 (479):\penalty0 824--840, September 2007.
\newblock ISSN 0162-1459.
\newblock \doi{10.1198/016214506000000780}.

\bibitem[Cressie and Wikle(2011)]{cressie_spatiotemporal:2011}
Noel A.~C. Cressie and Christopher~K. Wikle.
\newblock \emph{Statistics for Spatio-Temporal Data}.
\newblock Wiley, Hoboken, N.J., 2011.
\newblock ISBN 978-0-471-69274-4.

\bibitem[Davison et~al.(2012)Davison, Padoan, and Ribatet]{davison_spatialextremes:2012}
A.~C. Davison, S.~A. Padoan, and M.~Ribatet.
\newblock Statistical {{Modeling}} of {{Spatial Extremes}}.
\newblock \emph{Statistical Science}, 27\penalty0 (2), May 2012.
\newblock ISSN 0883-4237.
\newblock \doi{10.1214/11-STS376}.

\bibitem[Dittes et~al.(2018)Dittes, {\v S}pa{\v c}kov{\'a}, Schoppa, and Straub]{dittes_uncertainty:2018}
Beatrice Dittes, Olga {\v S}pa{\v c}kov{\'a}, Lukas Schoppa, and Daniel Straub.
\newblock Managing uncertainty in flood protection planning with climate projections.
\newblock \emph{Hydrology and Earth System Sciences}, 22\penalty0 (4):\penalty0 2511--2526, 2018.
\newblock \doi{10.5194/hess-22-2511-2018}.

\bibitem[Donat et~al.(2016)Donat, Lowry, Alexander, O'Gorman, and Maher]{donat_moreextreme:2016}
Markus~G. Donat, Andrew~L. Lowry, Lisa~V. Alexander, Paul~A. O'Gorman, and Nicola Maher.
\newblock More extreme precipitation in the world's dry and wet regions.
\newblock \emph{Nature Climate Change}, 6\penalty0 (5):\penalty0 508--513, May 2016.
\newblock ISSN 1758-6798.
\newblock \doi{10.1038/nclimate2941}.

\bibitem[{Doss-Gollin} et~al.(2019){Doss-Gollin}, Farnham, Steinschneider, and Lall]{doss-gollin_robustadaptation:2019}
James {Doss-Gollin}, David~J. Farnham, Scott Steinschneider, and Upmanu Lall.
\newblock Robust adaptation to multiscale climate variability.
\newblock \emph{Earth's Future}, 7\penalty0 (7):\penalty0 734--747, June 2019.
\newblock ISSN 2328-4277.
\newblock \doi{10.1029/2019ef001154}.

\bibitem[Dyrrdal et~al.(2015)Dyrrdal, Lenkoski, Thorarinsdottir, and Stordal]{dyrrdal_bayesrain:2015}
Anita~Verpe Dyrrdal, Alex Lenkoski, Thordis~L. Thorarinsdottir, and Frode Stordal.
\newblock Bayesian hierarchical modeling of extreme hourly precipitation in {{Norway}}.
\newblock \emph{Environmetrics}, 26\penalty0 (2):\penalty0 89--106, 2015.
\newblock ISSN 1099-095X.
\newblock \doi{10.1002/env.2301}.

\bibitem[Ehret et~al.(2012)Ehret, Zehe, Wulfmeyer, {Warrach-Sagi}, and Liebert]{ehret_biascorrection:2012}
U~Ehret, E~Zehe, V~Wulfmeyer, K~{Warrach-Sagi}, and J~Liebert.
\newblock Should we apply bias correction to global and regional climate model data?
\newblock \emph{Hydrology and Earth System Sciences}, 16\penalty0 (9):\penalty0 3391--3404, 2012.
\newblock \doi{10.5194/hess-16-3391-2012}.

\bibitem[Fagnant et~al.(2020)Fagnant, Gori, Sebastian, Bedient, and Ensor]{fagnant_spatiotemporal:2020}
Carlynn Fagnant, Avantika Gori, Antonia Sebastian, Philip~B. Bedient, and Katherine~B. Ensor.
\newblock Characterizing spatiotemporal trends in extreme precipitation in {{Southeast Texas}}.
\newblock \emph{Natural Hazards}, 104\penalty0 (2):\penalty0 1597--1621, November 2020.
\newblock ISSN 1573-0840.
\newblock \doi{10.1007/s11069-020-04235-x}.

\bibitem[Farnham et~al.(2018)Farnham, {Doss-Gollin}, and Lall]{farnham_credibly:2018}
David~J Farnham, James {Doss-Gollin}, and Upmanu Lall.
\newblock Regional extreme precipitation events: Robust inference from credibly simulated {{GCM}} variables.
\newblock \emph{Water Resources Research}, 54\penalty0 (6), 2018.
\newblock \doi{10.1002/2017wr021318}.

\bibitem[Feng et~al.(2019)Feng, Lian, Ying, Li, and Li]{feng_enso_diversity:2019}
Jie Feng, Tao Lian, Jun Ying, Junde Li, and Gen Li.
\newblock Do {{CMIP5}} models show {{El Ni{\~n}o}} diversity?
\newblock \emph{Journal of Climate}, November 2019.
\newblock ISSN 0894-8755.
\newblock \doi{10.1175/jcli-d-18-0854.1}.

\bibitem[Fowler and Kilsby(2003)]{fowler_rfauk:2003}
H.~J. Fowler and C.~G. Kilsby.
\newblock A regional frequency analysis of {{United Kingdom}} extreme rainfall from 1961 to 2000.
\newblock \emph{International Journal of Climatology}, 23\penalty0 (11):\penalty0 1313--1334, 2003.
\newblock ISSN 1097-0088.
\newblock \doi{10.1002/joc.943}.

\bibitem[Gavahi et~al.(2023)Gavahi, Foroumandi, and Moradkhani]{gavahi_precipfusion:2023}
Keyhan Gavahi, Ehsan Foroumandi, and Hamid Moradkhani.
\newblock A deep learning-based framework for multi-source precipitation fusion.
\newblock \emph{Remote Sensing of Environment}, 295:\penalty0 113723, September 2023.
\newblock ISSN 0034-4257.
\newblock \doi{10.1016/j.rse.2023.113723}.

\bibitem[Gelfand et~al.(2005)Gelfand, Banerjee, and Gamerman]{gelfand_spatial:2005}
Alan~E. Gelfand, Sudipto Banerjee, and Dani Gamerman.
\newblock Spatial process modelling for univariate and multivariate dynamic spatial data.
\newblock \emph{Environmetrics}, 16\penalty0 (5):\penalty0 465--479, August 2005.
\newblock ISSN 1180-4009, 1099-095X.
\newblock \doi{10.1002/env.715}.

\bibitem[Gelman et~al.(2014)Gelman, Carlin, Stern, and Rubin]{gelman_bda3:2014}
Andrew Gelman, John~B Carlin, Hal~S Stern, and Donald~B Rubin.
\newblock \emph{Bayesian {{Data Analysis}}}.
\newblock Chapman \& Hall/CRC Boca Raton, FL, USA, 3 edition, 2014.

\bibitem[Gelman et~al.(2020)Gelman, Vehtari, Simpson, Margossian, Carpenter, Yao, Kennedy, Gabry, B{\"u}rkner, and Modr{\'a}k]{gelman_workflow:2020}
Andrew Gelman, Aki Vehtari, Daniel Simpson, Charles~C. Margossian, Bob Carpenter, Yuling Yao, Lauren Kennedy, Jonah Gabry, Paul-Christian B{\"u}rkner, and Martin Modr{\'a}k.
\newblock Bayesian workflow.
\newblock \emph{arXiv:2011.01808 [stat]}, November 2020.
\newblock \doi{10.48550/arXiv.2011.01808}.

\bibitem[Goovaerts(2000)]{goovaerts_addelevation:2000}
P.~Goovaerts.
\newblock Geostatistical approaches for incorporating elevation into the spatial interpolation of rainfall.
\newblock \emph{Journal of Hydrology}, 228\penalty0 (1):\penalty0 113--129, February 2000.
\newblock ISSN 0022-1694.
\newblock \doi{10.1016/S0022-1694(00)00144-X}.

\bibitem[Gu et~al.(2022)Gu, Ye, Xin, Zhang, Zeng, Nerantzaki, and Papalexiou]{gu_extremechina:2022}
Xuezhi Gu, Lei Ye, Qian Xin, Chi Zhang, Fanzhang Zeng, Sofia~D. Nerantzaki, and Simon~Michael Papalexiou.
\newblock Extreme {{Precipitation}} in {{China}}: {{A Review}} on {{Statistical Methods}} and {{Applications}}.
\newblock \emph{Advances in Water Resources}, 163:\penalty0 104144, May 2022.
\newblock ISSN 03091708.
\newblock \doi{10.1016/j.advwatres.2022.104144}.

\bibitem[Haruna et~al.(2023)Haruna, Blanchet, and Favre]{haruna_comparison:2023}
Abubakar Haruna, Juliette Blanchet, and Anne-Catherine Favre.
\newblock Modeling {{Intensity}}-{{Duration}}-{{Frequency Curves}} for the {{Whole Range}} of {{Non}}-{{Zero Precipitation}}: {{A Comparison}} of {{Models}}.
\newblock \emph{Water Resources Research}, 59\penalty0 (6):\penalty0 e2022WR033362, June 2023.
\newblock ISSN 0043-1397, 1944-7973.
\newblock \doi{10.1029/2022WR033362}.

\bibitem[Hosking(1990)]{hosking_lmoments:1990}
J.~R.~M. Hosking.
\newblock L-{{Moments}}: {{Analysis}} and {{Estimation}} of {{Distributions Using Linear Combinations}} of {{Order Statistics}}.
\newblock \emph{Journal of the Royal Statistical Society. Series B (Methodological)}, 52\penalty0 (1):\penalty0 105--124, 1990.
\newblock ISSN 0035-9246.

\bibitem[Hosking(1997)]{hosking_rfa:1997}
J.~R.~M. Hosking.
\newblock \emph{Regional Frequency Analysis: An Approach Based on {{L-moments}} / {{J}}.{{R}}.{{M}}. {{Hosking}} and {{J}}.{{R}}. {{Wallis}}.}
\newblock Cambridge University Press, Cambridge, United Kingdom, 1997.
\newblock ISBN 978-0-521-43045-6.

\bibitem[Hu et~al.(2021)Hu, Yao, He, and Chen]{hu_elev:2021}
Wenfeng Hu, Junqiang Yao, Qing He, and Jing Chen.
\newblock Elevation-{{Dependent Trends}} in {{Precipitation Observed}} over and around the {{Tibetan Plateau}} from 1971 to 2017.
\newblock \emph{Water}, 13\penalty0 (20):\penalty0 2848, October 2021.
\newblock ISSN 2073-4441.
\newblock \doi{10.3390/w13202848}.

\bibitem[Jorgensen and {Nielsen-Gammon}(2024)]{jorgensen_nonstationary:2024}
Savannah~K. Jorgensen and John~W. {Nielsen-Gammon}.
\newblock Nonstationarity in {{Extreme Precipitation Return Values Along}} the {{United States Gulf}} and {{Southeastern Coasts}}.
\newblock \emph{Journal of Hydrometeorology}, -1\penalty0 (aop), March 2024.
\newblock ISSN 1525-7541, 1525-755X.
\newblock \doi{10.1175/JHM-D-22-0157.1}.

\bibitem[Kates et~al.(2006)Kates, Colten, Laska, and Leatherman]{kates_nolakatrina:2006}
R.~W. Kates, C.~E. Colten, S.~Laska, and S.~P. Leatherman.
\newblock Reconstruction of {{New Orleans}} after {{Hurricane Katrina}}: {{A}} research perspective.
\newblock \emph{Proceedings of the National Academy of Sciences}, 103\penalty0 (40):\penalty0 14653--14660, October 2006.
\newblock \doi{10.1073/pnas.0605726103}.

\bibitem[Katz et~al.(2002)Katz, Parlange, and Naveau]{katz_hydroextrems:2002}
R~W Katz, M~B Parlange, and P~Naveau.
\newblock Statistics of extremes in hydrology.
\newblock \emph{Advances in Water Resources}, 25\penalty0 (8-12):\penalty0 1287--1304, 2002.
\newblock \doi{10.1016/s0309-1708(02)00056-8}.

\bibitem[Keeling et~al.(1976)Keeling, Bacastow, Bainbridge, Ekdahl, Guenther, Waterman, and Chin]{keeling_maunaloa:1976}
Charles~D. Keeling, Robert~B. Bacastow, Arnold~E. Bainbridge, Carl~A. Ekdahl, Peter~R. Guenther, Lee~S. Waterman, and John F.~S. Chin.
\newblock Atmospheric carbon dioxide variations at {{Mauna Loa Observatory}}, {{Hawaii}}.
\newblock \emph{Tellus A: Dynamic Meteorology and Oceanography}, 28\penalty0 (6):\penalty0 538, January 1976.
\newblock ISSN 1600-0870.
\newblock \doi{10.3402/tellusa.v28i6.11322}.

\bibitem[Koenker and Machado(1999)]{koenker_goodnessfit:1999}
Roger Koenker and Jos{\'e} A.~F. Machado.
\newblock Goodness of {{Fit}} and {{Related Inference Processes}} for {{Quantile Regression}}.
\newblock \emph{Journal of the American Statistical Association}, 94\penalty0 (448):\penalty0 1296--1310, December 1999.
\newblock ISSN 0162-1459.
\newblock \doi{10.1080/01621459.1999.10473882}.

\bibitem[Kourtis and Tsihrintzis(2022)]{kourtis_IDFchange:2022}
Ioannis~M. Kourtis and Vassilios~A. Tsihrintzis.
\newblock Update of intensity-duration-frequency ({{IDF}}) curves under climate change: A review.
\newblock \emph{Water Supply}, 22\penalty0 (5):\penalty0 4951--4974, March 2022.
\newblock ISSN 1606-9749.
\newblock \doi{10.2166/ws.2022.152}.

\bibitem[Lafferty and Sriver(2023)]{lafferty_BCuncertain:2023}
David~C. Lafferty and Ryan~L. Sriver.
\newblock Downscaling and bias-correction contribute considerable uncertainty to local climate projections in {{CMIP6}}.
\newblock \emph{npj Climate and Atmospheric Science}, 6\penalty0 (1):\penalty0 1--13, September 2023.
\newblock ISSN 2397-3722.
\newblock \doi{10.1038/s41612-023-00486-0}.

\bibitem[Lall et~al.(2018)Lall, Johnson, Colohan, Aghakouchak, Arumugam, Brown, Mccabe, and Pulwarty]{lall_ncawater:2018}
Upmanu Lall, Thomas Johnson, Peter Colohan, Amir Aghakouchak, Sankar Arumugam, Casey Brown, Gregory~J. Mccabe, and Roger~S. Pulwarty.
\newblock \emph{Chapter 3: {{Water}}}.
\newblock U.S. Global Change Research Program, Washington, D.C., 2018.
\newblock \doi{10.7930/NCA4.2018.CH3}.

\bibitem[Lee and Haran(2022)]{lee_picar:2022}
Ben~Seiyon Lee and Murali Haran.
\newblock {{PICAR}}: {{An Efficient Extendable Approach}} for {{Fitting Hierarchical Spatial Models}}.
\newblock \emph{Technometrics}, 64\penalty0 (2):\penalty0 187--198, April 2022.
\newblock ISSN 0040-1706, 1537-2723.
\newblock \doi{10.1080/00401706.2021.1933596}.

\bibitem[Lima et~al.(2016)Lima, Lall, Troy, and Devineni]{lima_GEVflood:2016}
Carlos H~R Lima, Upmanu Lall, Tara Troy, and Naresh Devineni.
\newblock A hierarchical {{Bayesian GEV}} model for improving local and regional flood quantile estimates.
\newblock \emph{Journal of Hydrology}, 541:\penalty0 816--823, October 2016.
\newblock \doi{10.1016/j.jhydrol.2016.07.042}.

\bibitem[{Lopez-Cantu} and Samaras(2018)]{lopez-cantu_standards:2018}
Tania {Lopez-Cantu} and Constantine Samaras.
\newblock Temporal and spatial evaluation of stormwater engineering standards reveals risks and priorities across the {{United States}}.
\newblock \emph{Environmental Research Letters}, 13\penalty0 (7), June 2018.
\newblock ISSN 1748-9326.
\newblock \doi{10.1088/1748-9326/aac696}.

\bibitem[Luo et~al.(2023)Luo, Zhang, Yu, Liang, Xia, Gao, Gao, and Yin]{luo_zhengzhou2021:2023}
Yali Luo, Jiahua Zhang, Miao Yu, Xudong Liang, Rudi Xia, Yanyu Gao, Xiaoyu Gao, and Jinfang Yin.
\newblock On the {{Influences}} of {{Urbanization}} on the {{Extreme Rainfall}} over {{Zhengzhou}} on 20 {{July}} 2021: {{A Convection-Permitting Ensemble Modeling Study}}.
\newblock \emph{Advances in Atmospheric Sciences}, 40\penalty0 (3):\penalty0 393--409, March 2023.
\newblock ISSN 0256-1530, 1861-9533.
\newblock \doi{10.1007/s00376-022-2048-8}.

\bibitem[Mahmood et~al.(2014)Mahmood, Pielke, Hubbard, Niyogi, Dirmeyer, McAlpine, Carleton, Hale, Gameda, Beltr{\'a}n-Przekurat, Baker, McNider, Legates, Shepherd, Du, Blanken, Frauenfeld, Nair, and Fall]{mahmood_landcover:2014}
Rezaul Mahmood, Roger~A. Pielke, Kenneth~G. Hubbard, Dev Niyogi, Paul~A. Dirmeyer, Clive McAlpine, Andrew~M. Carleton, Robert Hale, Samuel Gameda, Adriana Beltr{\'a}n-Przekurat, Bruce Baker, Richard McNider, David~R. Legates, Marshall Shepherd, Jinyang Du, Peter~D. Blanken, Oliver~W. Frauenfeld, U.S. Nair, and Souleymane Fall.
\newblock Land cover changes and their biogeophysical effects on climate.
\newblock \emph{International Journal of Climatology}, 34\penalty0 (4):\penalty0 929--953, March 2014.
\newblock ISSN 0899-8418, 1097-0088.
\newblock \doi{10.1002/joc.3736}.

\bibitem[Martel et~al.(2021)Martel, Brissette, {Lucas-Picher}, Troin, and Arsenault]{martel_climatechange&IDF:2021}
Jean-Luc Martel, Fran{\c c}ois~P. Brissette, Philippe {Lucas-Picher}, Magali Troin, and Richard Arsenault.
\newblock Climate {{Change}} and {{Rainfall Intensity}}--{{Duration}}--{{Frequency Curves}}: {{Overview}} of {{Science}} and {{Guidelines}} for {{Adaptation}}.
\newblock \emph{Journal of Hydrologic Engineering}, 26\penalty0 (10):\penalty0 03121001, October 2021.
\newblock ISSN 1084-0699, 1943-5584.
\newblock \doi{10.1061/(ASCE)HE.1943-5584.0002122}.

\bibitem[Martins and Stedinger(2000)]{martins_estimators:2000}
Eduardo~S. Martins and Jery~R. Stedinger.
\newblock Generalized maximum-likelihood generalized extreme-value quantile estimators for hydrologic data.
\newblock \emph{Water Resources Research}, 36\penalty0 (3):\penalty0 737--744, 2000.
\newblock ISSN 1944-7973.
\newblock \doi{10.1029/1999WR900330}.

\bibitem[Merz et~al.(2014)Merz, Aerts, {Arnbjerg-Nielsen}, Baldi, Becker, Bichet, Bl{\"o}schl, Bouwer, Brauer, Cioffi, Delgado, Gocht, Guzzetti, Harrigan, Hirschboeck, Kilsby, Kron, Kwon, Lall, Merz, Nissen, Salvatti, Swierczynski, Ulbrich, Viglione, Ward, Weiler, Wilhelm, and Nied]{merz_review:2014}
Bruno Merz, Jeroen C J~H Aerts, Karsten {Arnbjerg-Nielsen}, M~Baldi, A~Becker, A~Bichet, G{\"u}nter Bl{\"o}schl, Laurens~M Bouwer, Achim Brauer, F~Cioffi, J~M Delgado, M~Gocht, F~Guzzetti, S~Harrigan, K~Hirschboeck, C~Kilsby, W~Kron, H~H Kwon, Upmanu Lall, R~Merz, K~Nissen, P~Salvatti, Tina Swierczynski, U~Ulbrich, A~Viglione, P~J Ward, M~Weiler, B~Wilhelm, and M~Nied.
\newblock Floods and climate: Emerging perspectives for flood risk assessment and management.
\newblock \emph{Natural Hazards and Earth System Science}, 14\penalty0 (7):\penalty0 1921--1942, 2014.
\newblock \doi{10/gb9nzm}.

\bibitem[Milly et~al.(2008)Milly, Betancourt, Falkenmark, Hirsch, Kundzewicz, Lettenmaier, and Stouffer]{milly_stationary:2008}
P~C~D Milly, Julio Betancourt, M~Falkenmark, R~M Hirsch, Z~W Kundzewicz, D~P Lettenmaier, and R~J Stouffer.
\newblock Stationarity is dead: Whither water management?
\newblock \emph{Science}, 319\penalty0 (5863):\penalty0 573--574, February 2008.
\newblock \doi{10.1126/science.1151915}.

\bibitem[Miniussi and Marani(2020)]{miniussi_metastatistical:2020}
Arianna Miniussi and Marco Marani.
\newblock Estimation of {{Daily Rainfall Extremes Through}} the {{Metastatistical Extreme Value Distribution}}: {{Uncertainty Minimization}} and {{Implications}} for {{Trend Detection}}.
\newblock \emph{Water Resources Research}, 56\penalty0 (7):\penalty0 e2019WR026535, 2020.
\newblock ISSN 1944-7973.
\newblock \doi{10.1029/2019WR026535}.

\bibitem[Mishra and Singh(2010)]{mishra_extremechange:2010}
Ashok~K. Mishra and Vijay~P. Singh.
\newblock Changes in extreme precipitation in {{Texas}}.
\newblock \emph{Journal of Geophysical Research: Atmospheres}, 115\penalty0 (D14):\penalty0 2009JD013398, July 2010.
\newblock ISSN 0148-0227.
\newblock \doi{10.1029/2009JD013398}.

\bibitem[Moftakhari et~al.(2018)Moftakhari, AghaKouchak, Sanders, Allaire, and Matthew]{moftakhari_nuisanceflooding:2018}
Hamed~R. Moftakhari, Amir AghaKouchak, Brett~F. Sanders, Maura Allaire, and Richard~A. Matthew.
\newblock What {{Is Nuisance Flooding}}? {{Defining}} and {{Monitoring}} an {{Emerging Challenge}}.
\newblock \emph{Water Resources Research}, 54\penalty0 (7):\penalty0 4218--4227, 2018.
\newblock ISSN 1944-7973.
\newblock \doi{10.1029/2018WR022828}.

\bibitem[Montanari and Koutsoyiannis(2014)]{montanari_immortal:2014}
Alberto Montanari and Demetris Koutsoyiannis.
\newblock Modeling and mitigating natural hazards: Stationarity is immortal!
\newblock \emph{Water Resources Research}, 50\penalty0 (12):\penalty0 9748--9756, December 2014.
\newblock \doi{10.1002/2014wr016092}.

\bibitem[{Nielsen-Gammon}(2020)]{nielsen-gammon_harris:2020}
John~W. {Nielsen-Gammon}.
\newblock Observation-based estimates of present-day and future climate change impacts on heavy rainfall in {{Harris County}}.
\newblock Technical report, June 2020.

\bibitem[O'Gorman(2015)]{ogorman_extremes:2015}
Paul~A O'Gorman.
\newblock Precipitation extremes under climate change.
\newblock \emph{Current Climate Change Reports}, 1\penalty0 (2):\penalty0 49--59, April 2015.
\newblock \doi{10.1007/s40641-015-0009-3}.

\bibitem[Ossand{\'o}n et~al.(2021)Ossand{\'o}n, Rajagopalan, and Kleiber]{ossandon_semibayesian:2021}
{\'A}lvaro Ossand{\'o}n, Balaji Rajagopalan, and William Kleiber.
\newblock Spatial-temporal multivariate semi-{{Bayesian}} hierarchical framework for extreme precipitation frequency analysis.
\newblock \emph{Journal of Hydrology}, 600:\penalty0 126499, September 2021.
\newblock ISSN 0022-1694.
\newblock \doi{10.1016/j.jhydrol.2021.126499}.

\bibitem[Pendergrass et~al.(2017)Pendergrass, Knutti, Lehner, Deser, and Sanderson]{pendergrass_variability:2017}
Angeline~G. Pendergrass, Reto Knutti, Flavio Lehner, Clara Deser, and Benjamin~M. Sanderson.
\newblock Precipitation variability increases in a warmer climate.
\newblock \emph{Scientific Reports}, 7\penalty0 (1):\penalty0 1--9, December 2017.
\newblock ISSN 2045-2322.
\newblock \doi{10.1038/s41598-017-17966-y}.

\bibitem[Perica et~al.(2013)Perica, Martin, Pavlovic, Roy, St.~Laurent, Trypaluk, Unruh, Yekta, and Bonnin]{perica_atlas14southeastern:2013}
Sanja Perica, Deborah Martin, Sandra Pavlovic, Ishani Roy, Michael St.~Laurent, Carl Trypaluk, Dale Unruh, Michael Yekta, and Geoffrey Bonnin.
\newblock {{NOAA Atlas}} 14.
\newblock Technical Report Volume 9 Version 2.0: Southeastern States (Alabama, Arkansas, Florida, Georgia, Louisiana, Mississippi), {National Weather Service, National Oceanic and Atmospheric Administration, U.S. Department of Commerce}, Silver Spring, MD, 2013.

\bibitem[Perica et~al.(2018)Perica, Pavlovic, St.~Laurent, Trypaluk, Unruh, and Wilhite]{perica_atlas14_texas:2018}
Sanja Perica, Sandra Pavlovic, Michael St.~Laurent, Carl Trypaluk, Dale Unruh, and Orlan Wilhite.
\newblock {{NOAA Atlas}} 14.
\newblock Technical Report Volume 11 Version 2.0: Texas, {National Weather Service, National Oceanic and Atmospheric Administration, U.S. Department of Commerce}, Silver Spring, MD, 2018.

\bibitem[Pielke~Sr. et~al.(2007)Pielke~Sr., Adegoke, {Beltr{\'a}n-Przekurat}, Hiemstra, Lin, Nair, Niyogi, and Nobis]{pielkesr_landimpact:2007}
R.~A. Pielke~Sr., J.~Adegoke, A.~{Beltr{\'a}n-Przekurat}, C.~A. Hiemstra, J.~Lin, U.~S. Nair, D.~Niyogi, and T.~E. Nobis.
\newblock An overview of regional land-use and land-cover impacts on rainfall.
\newblock \emph{Tellus B: Chemical and Physical Meteorology}, 59\penalty0 (3):\penalty0 587, January 2007.
\newblock ISSN 1600-0889, 0280-6509.
\newblock \doi{10.1111/j.1600-0889.2007.00251.x}.

\bibitem[Ragno et~al.(2019)Ragno, AghaKouchak, Cheng, and Sadegh]{ragno_processinformed:2019}
Elisa Ragno, Amir AghaKouchak, Linyin Cheng, and Mojtaba Sadegh.
\newblock A generalized framework for process-informed nonstationary extreme value analysis.
\newblock \emph{Advances in Water Resources}, 130:\penalty0 270--282, August 2019.
\newblock ISSN 0309-1708.
\newblock \doi{10.1016/j.advwatres.2019.06.007}.

\bibitem[Ramsay and Silverman(2005)]{ramsay_dataanalysis:2005}
J.~O. Ramsay and B.~W. Silverman.
\newblock \emph{Functional {{Data Analysis}}}.
\newblock Springer {{Series}} in {{Statistics}}. Springer New York, New York, NY, second edition edition, 2005.
\newblock ISBN 978-0-387-40080-8.
\newblock \doi{10.1007/b98888}.

\bibitem[Rasmussen and Williams(2006)]{rasmussen_gp4ml:2006}
Carl~Edward Rasmussen and Chris K~I Williams.
\newblock \emph{Gaussian {{Processes}} for Machine Learning}.
\newblock the MIT Press, 2006.
\newblock ISBN 0-262-18253-X.

\bibitem[Risser and Wehner(2017)]{risser_Harvey:2017}
Mark~D. Risser and Michael~F. Wehner.
\newblock Attributable {{Human}}-{{Induced Changes}} in the {{Likelihood}} and {{Magnitude}} of the {{Observed Extreme Precipitation}} during {{Hurricane Harvey}}.
\newblock \emph{Geophysical Research Letters}, 44\penalty0 (24), December 2017.
\newblock ISSN 0094-8276, 1944-8007.
\newblock \doi{10.1002/2017GL075888}.

\bibitem[Rosenzweig et~al.(2018)Rosenzweig, McPhillips, Chang, Cheng, Welty, Matsler, Iwaniec, and Davidson]{rosenzweig_pluvial:2018}
Bernice~R. Rosenzweig, Lauren McPhillips, Heejun Chang, Chingwen Cheng, Claire Welty, Marissa Matsler, David Iwaniec, and Cliff~I. Davidson.
\newblock Pluvial flood risk and opportunities for resilience.
\newblock \emph{WIREs Water}, 5\penalty0 (6), November 2018.
\newblock ISSN 2049-1948, 2049-1948.
\newblock \doi{10.1002/wat2.1302}.

\bibitem[Rubino et~al.(2019)Rubino, Etheridge, Thornton, Allison, Francey, Langenfelds, Steele, Trudinger, Spencer, Curran, Van~Ommen, and Smith]{rubino_lawdome:2019}
Mauro Rubino, David Etheridge, David Thornton, Colin Allison, Roger Francey, Ray Langenfelds, Paul Steele, Cathy Trudinger, Darren Spencer, Mark Curran, Tas Van~Ommen, and Andrew Smith.
\newblock Law {{Dome Ice Core}} 2000-{{Year CO2}}, {{CH4}}, {{N2O}} and {{d13C-CO2}}, 2019.

\bibitem[Russell et~al.(2020)Russell, Risser, Smith, and Kunkel]{russell_gmxsst:2020}
Brook~T. Russell, Mark~D. Risser, Richard~L. Smith, and Kenneth~E. Kunkel.
\newblock Investigating the association between late spring {{Gulf}} of {{Mexico}} sea surface temperatures and {{U}}.{{S}}. {{Gulf Coast}} precipitation extremes with focus on {{Hurricane Harvey}}.
\newblock \emph{Environmetrics}, 31\penalty0 (2):\penalty0 e2595, 2020.
\newblock ISSN 1099-095X.
\newblock \doi{10.1002/env.2595}.

\bibitem[Salas et~al.(2018)Salas, Obeysekera, and Vogel]{salas_review:2018}
J~D Salas, J~Obeysekera, and R~M Vogel.
\newblock Techniques for assessing water infrastructure for nonstationary extreme events: A review.
\newblock \emph{Hydrological Sciences Journal}, 63\penalty0 (3):\penalty0 325--352, 2018.
\newblock \doi{10.1080/02626667.2018.1426858}.

\bibitem[Schlef et~al.(2023)Schlef, Kunkel, Brown, Demissie, Lettenmaier, Wagner, Wigmosta, Karl, Easterling, Wang, Fran{\c c}ois, and Yan]{schlef_idf:2023}
Katherine~E. Schlef, Kenneth~E. Kunkel, Casey Brown, Yonas Demissie, Dennis~P. Lettenmaier, Anna Wagner, Mark~S. Wigmosta, Thomas~R. Karl, David~R. Easterling, Kimberly~J. Wang, Baptiste Fran{\c c}ois, and Eugene Yan.
\newblock Incorporating non-stationarity from climate change into rainfall frequency and intensity-duration-frequency ({{IDF}}) curves.
\newblock \emph{Journal of Hydrology}, 616:\penalty0 128757, January 2023.
\newblock ISSN 0022-1694.
\newblock \doi{10.1016/j.jhydrol.2022.128757}.

\bibitem[Schmidt and Gelfand(2003)]{schmidt_coregionalization:2003}
Alexandra~M. Schmidt and Alan~E. Gelfand.
\newblock A {{Bayesian}} coregionalization approach for multivariate pollutant data.
\newblock \emph{Journal of Geophysical Research: Atmospheres}, 108\penalty0 (D24):\penalty0 2002JD002905, December 2003.
\newblock ISSN 0148-0227.
\newblock \doi{10.1029/2002JD002905}.

\bibitem[Selten(1998)]{selten_quadraticscoring:1998}
Reinhard Selten.
\newblock Axiomatic {{Characterization}} of the {{Quadratic Scoring Rule}}.
\newblock \emph{Experimental Economics}, 1\penalty0 (1):\penalty0 43--61, June 1998.
\newblock ISSN 1573-6938.
\newblock \doi{10.1023/A:1009957816843}.

\bibitem[Seneviratne et~al.(2021)Seneviratne, Zhang, Adnan, Badi, Dereczynski, Di~Luca, Ghosh, Iskandar, Kossin, Lewis, Otto, Pinto, Satoh, {Vicente-Serrano}, Wehner, and Zhou]{seneviratne_IPCC:2021}
S.I. Seneviratne, X.~Zhang, M.~Adnan, W.~Badi, C.~Dereczynski, A.~Di~Luca, S.~Ghosh, I.~Iskandar, J.~Kossin, S.~Lewis, F.~Otto, I.~Pinto, M.~Satoh, S.M. {Vicente-Serrano}, M.~Wehner, and B.~Zhou.
\newblock Weather and climate extreme events in a changing climate.
\newblock In V.~{Masson-Delmotte}, P.~Zhai, A.~Pirani, S.~L. Connors, C.~P{\'e}an, S.~Berger, N.~Caud, Y.~Chen, L.~Goldfarb, M.~I. Gomis, M.~Huang, K.~Leitzell, E.~Lonnoy, J.~B.~R. Matthews, T.~K. Maycock, T.~Waterfield, O.~Yelek{\c c}i, R.~Yu, and B.~Zhou, editors, \emph{Climate Change 2021: {{The}} Physical Science Basis. {{Contribution}} of Working Group {{I}} to the Sixth Assessment Report of the Intergovernmental Panel on Climate Change}, book section~11. Cambridge University Press, Cambridge, UK and New York, NY, USA, 2021.
\newblock \doi{10.1017/9781009157896.013}.

\bibitem[Serinaldi and Kilsby(2015)]{serinaldi_undead:2015}
Francesco Serinaldi and Chris~G Kilsby.
\newblock Stationarity is undead: Uncertainty dominates the distribution of extremes.
\newblock \emph{Advances in Water Resources}, 77:\penalty0 17--36, March 2015.
\newblock \doi{10.1016/j.advwatres.2014.12.013}.

\bibitem[Sharma et~al.(2021)Sharma, Lee, Nicholas, and Keller]{sharma_stormwater:2021}
Sanjib Sharma, Ben~Seiyon Lee, Robert~E. Nicholas, and Klaus Keller.
\newblock A {{Safety Factor Approach}} to {{Designing Urban Infrastructure}} for {{Dynamic Conditions}}.
\newblock \emph{Earth's Future}, 9\penalty0 (12):\penalty0 e2021EF002118, 2021.
\newblock ISSN 2328-4277.
\newblock \doi{10.1029/2021EF002118}.

\bibitem[Silva et~al.(2021)Silva, Simonovic, Schardong, and Goldenfum]{silva_can_idf:2021}
Daniele~Feitoza Silva, Slobodan~P. Simonovic, Andre Schardong, and Joel~Avruch Goldenfum.
\newblock Assessment of non-stationary {{IDF}} curves under a changing climate: {{Case}} study of different climatic zones in {{Canada}}.
\newblock \emph{Journal of Hydrology: Regional Studies}, 36:\penalty0 100870, August 2021.
\newblock ISSN 2214-5818.
\newblock \doi{10.1016/j.ejrh.2021.100870}.

\bibitem[Sobel et~al.(2023)Sobel, Lee, Bowen, Camargo, Cane, Clement, Fosu, Hart, Reed, Seager, and Tippett]{sobel_biases:2023}
Adam~H. Sobel, Chia-Ying Lee, Steven~G. Bowen, Suzana~J. Camargo, Mark~A. Cane, Amy Clement, Boniface Fosu, Megan Hart, Kevin~A. Reed, Richard Seager, and Michael~K. Tippett.
\newblock Near-term tropical cyclone risk and coupled {{Earth}} system model biases.
\newblock \emph{Proceedings of the National Academy of Sciences}, 120\penalty0 (33):\penalty0 e2209631120, August 2023.
\newblock \doi{10.1073/pnas.2209631120}.

\bibitem[Song et~al.(2020)Song, Zou, Mo, Zhang, Zhang, and Tian]{song_extremechina:2020}
Xiaomeng Song, Xianju Zou, Yuchen Mo, Jianyun Zhang, Chunhua Zhang, and Yimin Tian.
\newblock Nonstationary bayesian modeling of precipitation extremes in the {{Beijing-Tianjin-Hebei Region}}, {{China}}.
\newblock \emph{Atmospheric Research}, 242:\penalty0 105006, September 2020.
\newblock ISSN 0169-8095.
\newblock \doi{10.1016/j.atmosres.2020.105006}.

\bibitem[{Stan Development Team}(2022)]{stan_userguide:2022}
{Stan Development Team}.
\newblock \emph{Stan User's Guide}.
\newblock Version 2.30 edition, 2022.

\bibitem[Statkewicz et~al.(2021)Statkewicz, Talbot, and Rappenglueck]{statkewicz_raintx:2021}
Madeline~D. Statkewicz, Robert Talbot, and Bernhard Rappenglueck.
\newblock Changes in precipitation patterns in {{Houston}}, {{Texas}}.
\newblock \emph{Environmental Advances}, 5:\penalty0 100073, October 2021.
\newblock ISSN 2666-7657.
\newblock \doi{10.1016/j.envadv.2021.100073}.

\bibitem[Stedinger(1997)]{stedinger_estimators:1997}
Jery~R. Stedinger.
\newblock Expected probability and annual damage estimators.
\newblock \emph{Journal of Water Resources Planning and Management}, 123\penalty0 (2):\penalty0 125--135, March 1997.
\newblock ISSN 0733-9496.
\newblock \doi{10.1061/(ASCE)0733-9496(1997)123:2(125)}.

\bibitem[Stephenson et~al.(2016)Stephenson, Lehmann, and Phatak]{stephenson_maxstable:2016}
Alec~G. Stephenson, Eric~A. Lehmann, and Aloke Phatak.
\newblock A max-stable process model for rainfall extremes at different accumulation durations.
\newblock \emph{Weather and Climate Extremes}, 13:\penalty0 44--53, September 2016.
\newblock ISSN 22120947.
\newblock \doi{10.1016/j.wace.2016.07.002}.

\bibitem[Sui et~al.(2024)Sui, Yang, Shepherd, and Niyogi]{sui_urban_prcp:2024}
Xinxin Sui, Zong-Liang Yang, Marshall Shepherd, and Dev Niyogi.
\newblock Global scale assessment of urban precipitation anomalies.
\newblock \emph{Proceedings of the National Academy of Sciences}, 121\penalty0 (38):\penalty0 e2311496121, September 2024.
\newblock \doi{10.1073/pnas.2311496121}.

\bibitem[Sun et~al.(2023)Sun, Fu, Wang, Zhang, Chen, Su, Wang, Tang, and Ma]{sun_henan2021:2023}
Jianhua Sun, Shenming Fu, Huijie Wang, Yuanchun Zhang, Yun Chen, Aifang Su, Yaqiang Wang, Huan Tang, and Ruoyun Ma.
\newblock Primary characteristics of the extreme heavy rainfall event over {{Henan}} in {{July}} 2021.
\newblock \emph{Atmospheric Science Letters}, 24\penalty0 (1):\penalty0 e1131, 2023.
\newblock ISSN 1530-261X.
\newblock \doi{10.1002/asl.1131}.

\bibitem[Sun et~al.(2021)Sun, Zhang, Zwiers, Westra, and Alexander]{sun_change:2021}
Qiaohong Sun, Xuebin Zhang, Francis Zwiers, Seth Westra, and Lisa~V. Alexander.
\newblock A {{Global}}, {{Continental}}, and {{Regional Analysis}} of {{Changes}} in {{Extreme Precipitation}}.
\newblock \emph{Journal of Climate}, 34\penalty0 (1):\penalty0 243--258, January 2021.
\newblock ISSN 0894-8755, 1520-0442.
\newblock \doi{10.1175/JCLI-D-19-0892.1}.

\bibitem[Talchabhadel et~al.(2018)Talchabhadel, Karki, Thapa, Maharjan, and Parajuli]{talchabhadel_spatiotemporal:2018}
Rocky Talchabhadel, Ramchandra Karki, Bhesh~Raj Thapa, Manisha Maharjan, and Binod Parajuli.
\newblock Spatio-temporal variability of extreme precipitation in {{Nepal}}.
\newblock \emph{International Journal of Climatology}, 38\penalty0 (11):\penalty0 4296--4313, September 2018.
\newblock ISSN 0899-8418, 1097-0088.
\newblock \doi{10.1002/joc.5669}.

\bibitem[Tedesco et~al.(2020)Tedesco, McAlpine, and Porter]{tedesco_exposure:2020}
Marco Tedesco, Steven McAlpine, and Jeremy~R. Porter.
\newblock Exposure of real estate properties to the 2018 {{Hurricane Florence}} flooding.
\newblock \emph{Natural Hazards and Earth System Sciences}, 20\penalty0 (3):\penalty0 907--920, April 2020.
\newblock ISSN 1561-8633.
\newblock \doi{10.5194/nhess-20-907-2020}.

\bibitem[{The Federal Emergency Management Agency}(2019)]{fema_hydrologic:2019}
{The Federal Emergency Management Agency}.
\newblock Guidance for flood risk analysis and mapping: Hydrology rainfall-runoff analysis.
\newblock Guidance {{Document}}~91, 2019.

\bibitem[Thiemann et~al.(2001)Thiemann, Trosset, Gupta, and Sorooshian]{thiemann_hydrobayes:2001}
M.~Thiemann, M.~Trosset, H.~Gupta, and S.~Sorooshian.
\newblock Bayesian recursive parameter estimation for hydrologic models.
\newblock \emph{Water Resources Research}, 37\penalty0 (10):\penalty0 2521--2535, 2001.
\newblock ISSN 1944-7973.
\newblock \doi{10.1029/2000WR900405}.

\bibitem[Ulrich et~al.(2020)Ulrich, Jurado, Peter, Scheibel, and Rust]{ulrich_idf:2020}
Jana Ulrich, Oscar~E. Jurado, Madlen Peter, Marc Scheibel, and Henning~W. Rust.
\newblock Estimating {{IDF Curves Consistently}} over {{Durations}} with {{Spatial Covariates}}.
\newblock \emph{Water}, 12\penalty0 (11):\penalty0 3119, November 2020.
\newblock ISSN 2073-4441.
\newblock \doi{10.3390/w12113119}.

\bibitem[{van der Wiel} et~al.(2017){van der Wiel}, Kapnick, {van Oldenborgh}, Whan, Philip, Vecchi, Singh, Arrighi, and Cullen]{vanderwiel_nola2016:2017}
Karin {van der Wiel}, Sarah~B. Kapnick, Geert~Jan {van Oldenborgh}, Kirien Whan, Sjoukje Philip, Gabriel~A. Vecchi, Roop~K. Singh, Julie Arrighi, and Heidi Cullen.
\newblock Rapid attribution of the {{August}} 2016 flood-inducing extreme precipitation in south {{Louisiana}} to climate change.
\newblock \emph{Hydrology and Earth System Sciences}, 21\penalty0 (2):\penalty0 897--921, February 2017.
\newblock ISSN 1027-5606.
\newblock \doi{10.5194/hess-21-897-2017}.

\bibitem[van Oldenborgh et~al.(2017)van Oldenborgh, van~der Wiel, Sebastian, Singh, Arrighi, Otto, Haustein, Li, Vecchi, and Cullen]{oldenborgh_attribution:2017}
Geert~Jan van Oldenborgh, Karin van~der Wiel, Antonia Sebastian, Roop Singh, Julie Arrighi, Friederike Otto, Karsten Haustein, Sihan Li, Gabriel Vecchi, and Heidi Cullen.
\newblock Attribution of extreme rainfall from {{Hurricane Harvey}}, {{August}} 2017.
\newblock \emph{Environmental Research Letters}, 12\penalty0 (12):\penalty0 124009, December 2017.
\newblock ISSN 1748-9326.
\newblock \doi{10.1088/1748-9326/aa9ef2}.

\bibitem[Wikle(2019)]{wikle_hierarchical:2019}
Christopher~K. Wikle.
\newblock Comparison of {{Deep Neural Networks}} and {{Deep Hierarchical Models}} for {{Spatio-Temporal Data}}.
\newblock \emph{arXiv:1902.08321 [cs, stat]}, February 2019.

\bibitem[Wright et~al.(2020)Wright, Yu, and England]{wright_transposition:2020}
Daniel~B. Wright, Guo Yu, and John~F. England.
\newblock Six decades of rainfall and flood frequency analysis using stochastic storm transposition: {{Review}}, progress, and prospects.
\newblock \emph{Journal of Hydrology}, 585:\penalty0 124816, June 2020.
\newblock ISSN 0022-1694.
\newblock \doi{10.1016/j.jhydrol.2020.124816}.

\bibitem[Zhang et~al.(2018)Zhang, Villarini, Vecchi, and Smith]{zhang_harvey:2018}
Wei Zhang, Gabriele Villarini, Gabriel~A. Vecchi, and James~A. Smith.
\newblock Urbanization exacerbated the rainfall and flooding caused by hurricane {{Harvey}} in {{Houston}}.
\newblock \emph{Nature}, 563\penalty0 (7731):\penalty0 384--388, November 2018.
\newblock ISSN 1476-4687.
\newblock \doi{10.1038/s41586-018-0676-z}.

\end{thebibliography}

\newpage
\appendix
\renewcommand{\thefigure}{A\arabic{figure}} % Set figure numbering to A1, A2, ...
\setcounter{figure}{0} % Reset figure counter to 0 for the appendix

\section{Supplement figures}

\begin{figure}[h]
    \centering
    \includegraphics[width=0.8\textwidth]{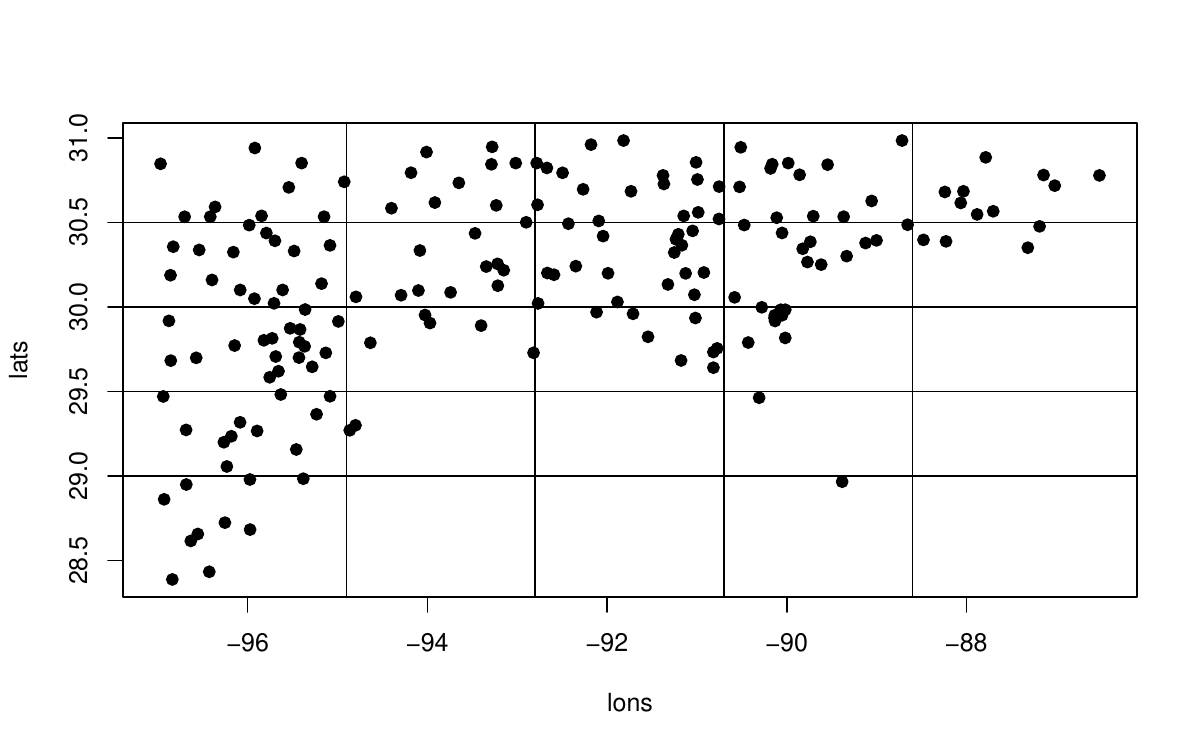}
    \caption{
        The study area is divided into \num{25} grids, which are utilized to create cross-validation subsets.
    }
    \label{fig:S1_Station_subsets}
\end{figure}

\begin{figure}[h]
    \centering
    \includegraphics[width=1\textwidth]{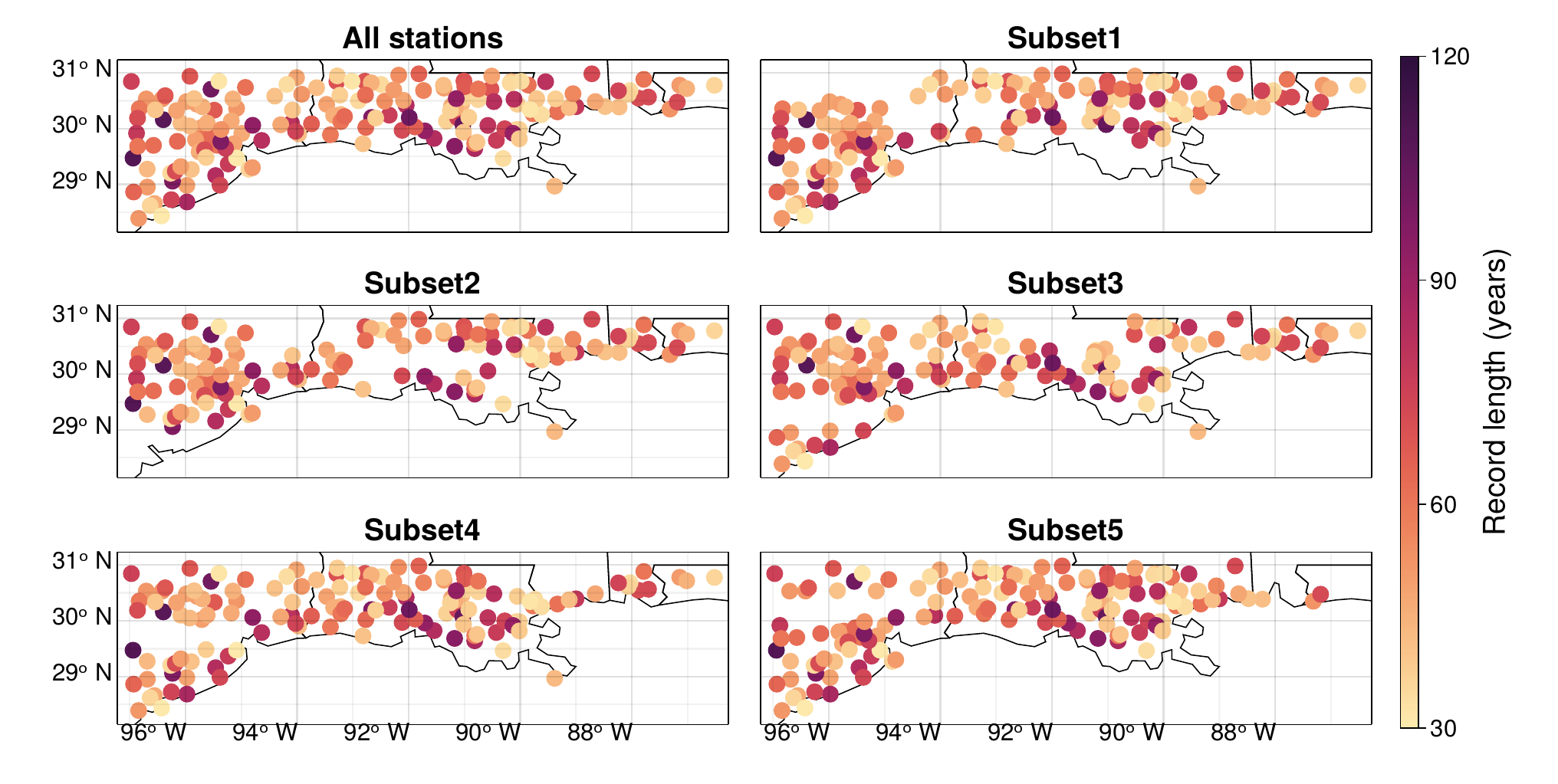}
    \caption{
        Maps show the full dataset and station subsets used for cross-validation along with the record length for each station.
    }
    \label{fig:S2_raw_data_nobs_subs}
\end{figure}

\begin{figure}[h]
    \centering
    \includegraphics[width=1\textwidth]{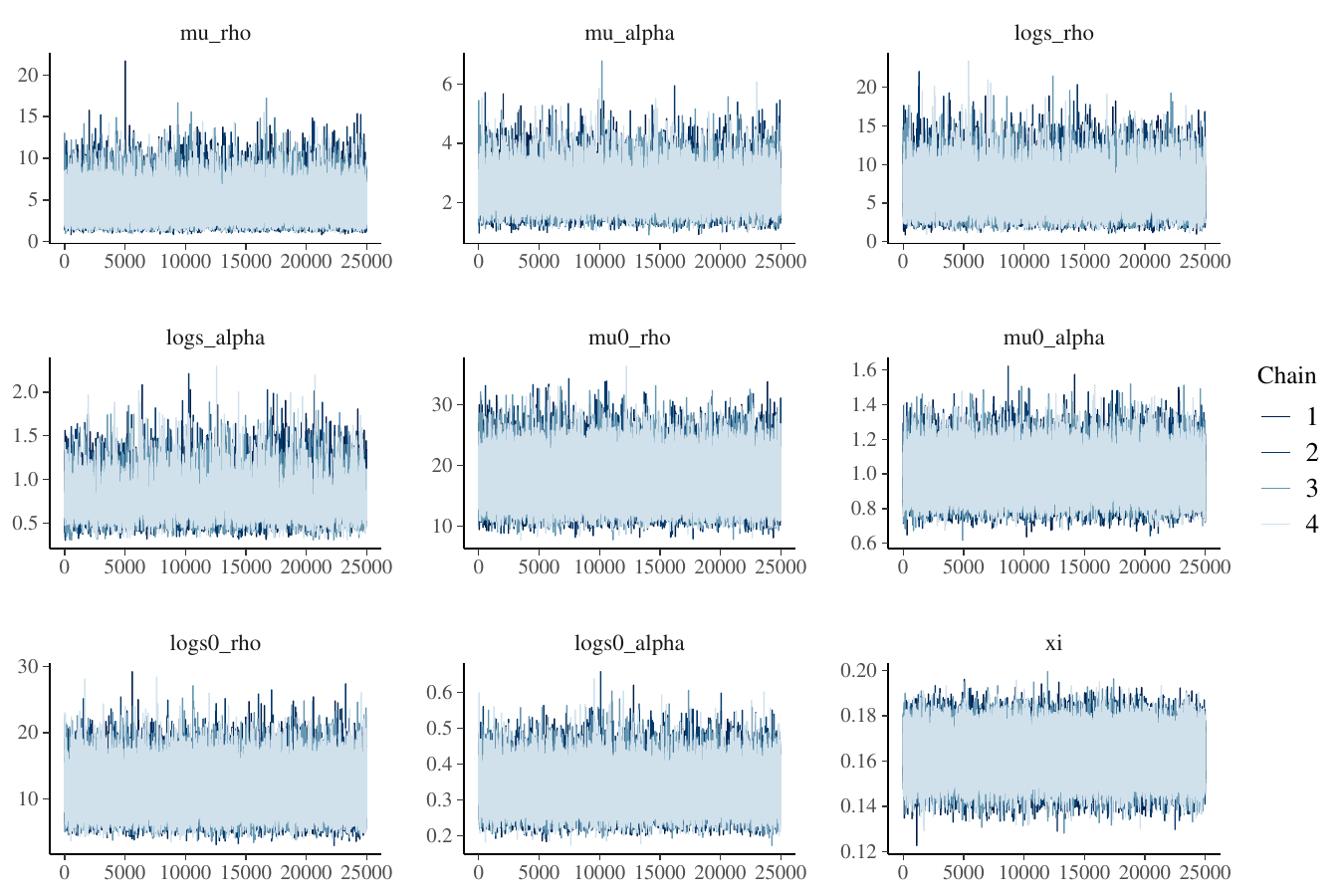}
    \caption{
        Trace plots indicate generally good posterior convergence.
        We present the trace plots for the key parameters obtained from \gls{mcmc} simulations using the full dataset.
    }
    \label{fig:S3_traceplots}
\end{figure}

\begin{figure}[h]
    \centering
    \includegraphics[width=1\textwidth]{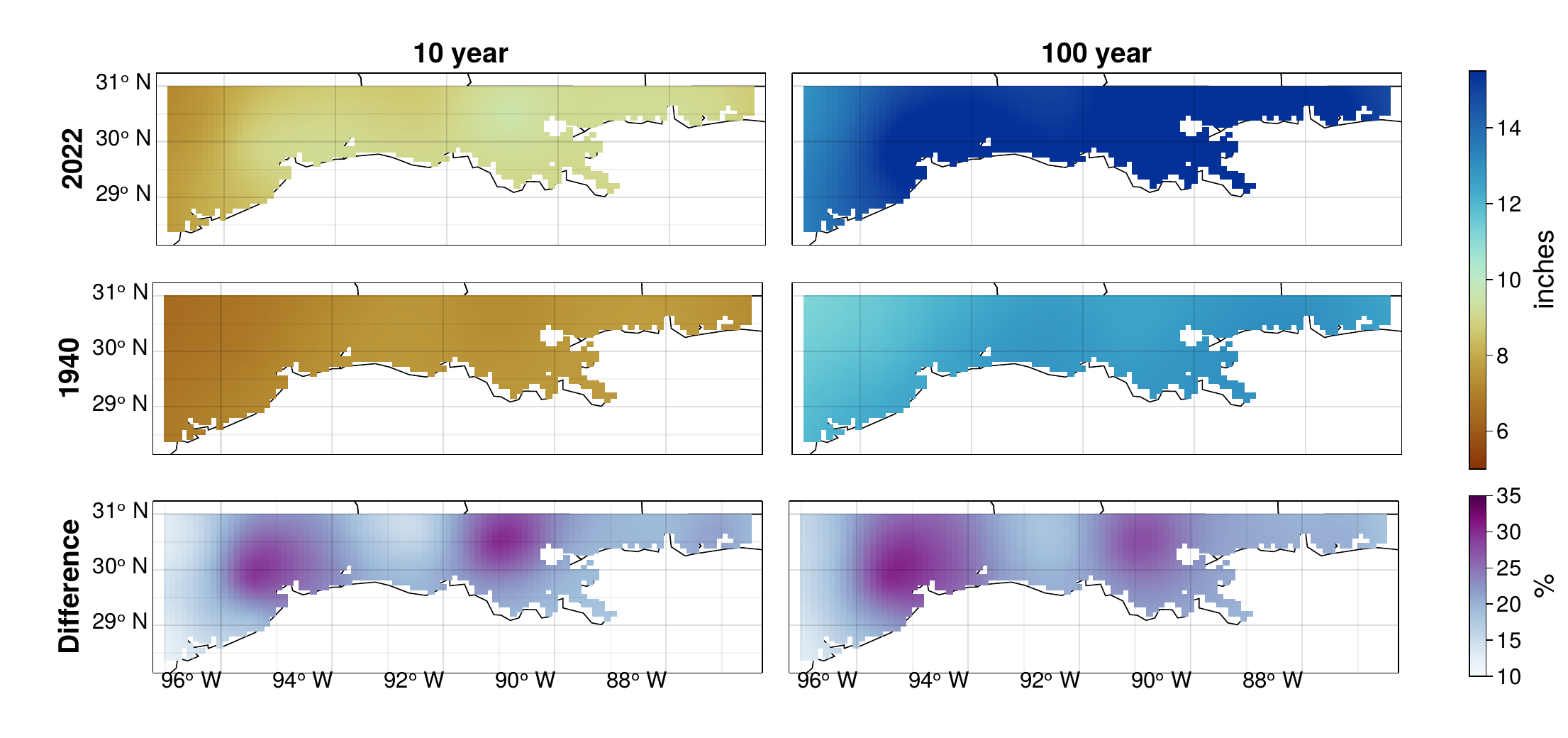}
    \caption{
        Similar to \cref{fig:spatio-temporal-patterns}, this figure presents the results in a gridded format. 
        It displays the gridded posterior mean of extreme precipitation return level estimates for the (L) \num{10}-year and (R) \num{100}-year events. The estimates are shown for (T) $2022$, (M) $1940$, and (B) the percentage change from $1940$ to $2022$. 
        These results are obtained using the Spatially Varying Covariates Model, where distribution parameters are interpolated with \gls{gp}.
    }
    \label{fig:S4_gridded}
\end{figure}

\end{document}